# Anisotropic In-plane Thermal Conductivity of Freestanding Few-layer $ReS_2$

Manavendra Pratap Singh[1], Nihal S. Kumar[2], Subhendu Mishra[2], Abhishek Jangid[3], Akshay Singh[3], Abhishek Kumar Singh[2], Akshay Naik[1*]

[1]*Centre for Nano Science and Engineering, Indian Institute of Science, Bengaluru, Karnataka 560012, India*

[2]*Materials Research Centre, Indian Institute of Science, Bengaluru, Karnataka 560012, India*

[3]*Department of Physics, Indian Institute of Science, Bengaluru, Karnataka 560012, India*

**Corresponding author: anaik@iisc.ac.in*

**Abstract:**

Rhenium disulfide ($ReS_2$) is a low-symmetry transition metal dichalcogenide (TMDC) exhibiting strong in-plane anisotropy, weak interlayer coupling, and stacking-dependent physical properties. While anisotropic thermal conductivity has been reported in bulk $ReS_2$, experimental studies on stacking-dependent thermal conductivity and its thickness evolution in the few-layer regime remain largely unexplored. Here, we have extracted the thermal conductivity of freestanding, few-layer $ReS_2$ samples (thickness < 10 nm) using polarization-resolved optothermal Raman thermometry after correcting for polarization dependent absorbance. All measured $ReS_2$ samples show pronounced in-plane anisotropic thermal conductivity. Notably, the ~3.5 nm AA-stacked flake shows higher thermal conductivity than the AB-stacked flake of the same thickness, highlighting the influence of stacking order on phonon transport. The in-plane thermal conductivity displays a non-monotonic dependence on thickness over the 2.5 to 8 nm range which is supported by density functional theory (DFT) calculations. These findings provide key insight into anisotropic phonon transport in low-symmetry 2D materials and highlight the potential of few-layer $ReS_2$ for nanoscale thermal management and thermoelectric applications.

**Keywords:** Freestanding $ReS_2$, Polarization-dependent Raman spectroscopy, Anisotropic thermal conductivity.

**Introduction:**

Transition-metal dichalcogenides (TMDCs) exhibit remarkable optoelectronic and thermal properties, making them promising materials for nanoscale device applications[1–5]. Conventional layered TMDCs such as $MX_2$ (M = Mo, W; X = S, Se) possess strong interlayer coupling, which leads to significant changes in their electronic band structure with thickness, including an indirect-to-direct bandgap transition in the monolayer limit. Owing to their tunable electronic and optical properties, these materials have been widely explored for piezotronics and optoelectronic applications[6–12]. In contrast, $ReS_2$ belongs to a distinct class of TMDCs with weak interlayer coupling due to a Peierls distortion of the octahedral 1T structure, resulting in a distorted 1T (1T′) structure[13–15]. Because of this distorted low-symmetry structure, $ReS_2$ retains a direct bandgap of ~1.55 eV largely independent of layer thickness[13], unlike conventional TMDCs. In addition, $ReS_2$ exhibits two stable stacking configurations, namely AA and AB stacking, which are known to influence its optical and vibrational properties[16–19].

Previous studies have reported strong in-plane anisotropy, resulting in directional optical and electrical properties of bulk and few-layer $ReS_2$[20–23]. Other anisotropic materials, including Td-$WTe_2$ and Black Phosphorus (BP), show significant in-plane variations[24,25]. Td-$WTe_2$ have thermal conductivity of 0.743 and 0.639 W/m-K along the zigzag and armchair axes, respectively[24]. Zhe *et al.* reported the impact of anisotropic phonon dispersion on thermal conductivities, with values of approximately 10 W/m-K and 20 W/m-K for zigzag and armchair configurations, respectively, in a ~10 nm thick BP flake[25]. Similar anisotropic behaviours have also been detected in the thermal studies of bulk $ReS_2$[26,27].

However, only a limited number of experimental studies have been conducted on the thermal conductivity of $ReS_2$. Jang et al. determined the orientation-dependent thermal conductivity of the exfoliated $ReS_2$ flakes with 60-450 nm thicknesses using time-domain thermoreflectance (TDTR)[26]. The in-plane thermal conductivity is higher along the *b*-axis than along the cross *b*-axis. Recently, Tahbaz et al. also reported the anisotropic thermal conductivity of $ReS_2$ using frequency domain thermoreflectance (FDTR) for a flake thickness of ~3 μm[27]. Both studies measured $ReS_2$ on substrate-supported configurations, where heat dissipation through the interface could potentially contribute to the apparent thermal transport. This impact of the substrate on the extracted thermal conductivity of $ReS_2$ across the thickness range was not explored as part of their work. An extensive study on the thermal conductivity of freestanding $ReS_2$, particularly focusing on stacking-dependent anisotropic thermal conductivity and its evolution for thicknesses below the 10 nm threshold, is still lacking. Such studies are crucial for the development of nanoscale devices for thermal transport, isolation, and anisotropic heat management.

In this work, we utilize polarization-resolved optothermal Raman technique to investigate anisotropic thermal conductivity of mechanically exfoliated freestanding $ReS_2$ flakes with thicknesses ranging from ~2.5 to 8 nm. To identify the crystallographic *b*-axis of the samples, we examined the polarization dependence of the Raman Spectra. We also performed white-light reflection measurements with incident polarization parallel and perpendicular to the *b*-axis to determine the optical absorbance required for accurate thermal conductivity estimation. Temperature and power-dependent Raman measurements performed with incident laser polarization ($I_{in}$) parallel and perpendicular to the *b*-axis and corrected for polarization dependent absorbance reveal pronounced in-plane anisotropic thermal conductivity. We show that thermal transport in $ReS_2$ strongly depends on both stacking order and thickness. AA-stacked sample indicate higher thermal conductivity than AB-stacked samples of the same thickness (3.5 nm), highlighting the role of interlayer interactions in phonon transport. Further, the thermal conductivity exhibits an apparent non-monotonic dependence on thickness over the ~2.5-8 nm range, indicating significant evolution of phonon dispersion and scattering mechanisms in few-layer $ReS_2$. These experimental observations are further supported by density functional theory (DFT) calculations. The results presented in this study are useful for the design and development of novel devices that leverage the anisotropic thermal properties of $ReS_2$.

**Results and discussion:**

An array of holes is created on 290 nm $SiO_2$ on top of a silicon wafer using optical lithography and reactive ion etching (RIE) (see Supporting Information S1 for more details). Commercially procured few-layer $ReS_2$ flakes were mechanically exfoliated onto polydimethyl siloxane (PDMS) and then transferred onto the array of holes fabricated on a $SiO_2$/Si substrate via a dry transfer technique[28] (see Methods). Figure 1b shows an optical image of the $ReS_2$ flake on the "holey" substrate of $SiO_2$/Si. The thickness of the transferred flakes was measured using atomic force microscopy (AFM) at the edge of the flake (see Figure 1c).

Figure 1d shows a simplified schematic of the experimental setup for optothermal Raman measurements. Figure 1e compares the Raman spectra of the $ReS_2$ flake acquired from the supported (on the substrate) and freestanding regions highlighted in Figure 1b. All the previously reported prominent Raman-active vibrational modes of $ReS_2$ are observed[14]. Here, we show only the first six Raman modes from 120 to 240 $cm^{-1}$, labeled as I, II, III, IV, V, and VI. The extended Raman spectra is presented in Supporting Information S2. There are prominent differences between the spectra observed for $ReS_2$ on the substrate and holes. There is a red shift for all the Raman modes when the measurements are performed on $ReS_2$ on holes, compared to when performed on the substrate. This redshift has previously been attributed to tensile strain[29,30 31]. The intensities of all the Raman mode peaks for the freestanding region are higher than those for the supported region. Additionally, the photoluminescence (PL) intensity of $ReS_2$ (Figure 1f) acquired from the hole region was ~6 times greater than that of the $SiO_2$-supported region. The Raman and PL intensity enhancements on the holes further confirm the freestanding nature[32,33].

$ReS_2$ is known to have two different stacking orders, which are likely to influence its thermal conductivity. To identify whether the stacking order is AA or AB, we use the separation between Raman modes I and III[16,17,34]. For monolayer $ReS_2$, the peak separation was ~ 17 $cm^{-1}$, smaller than the value typically reported for AB stacking and larger than that of AA stacking[16,17]. The separation between Raman modes III and I of the flake is 14.02 $cm^{-1}$, which confirms the AA stacking order (see Supporting Information S3).

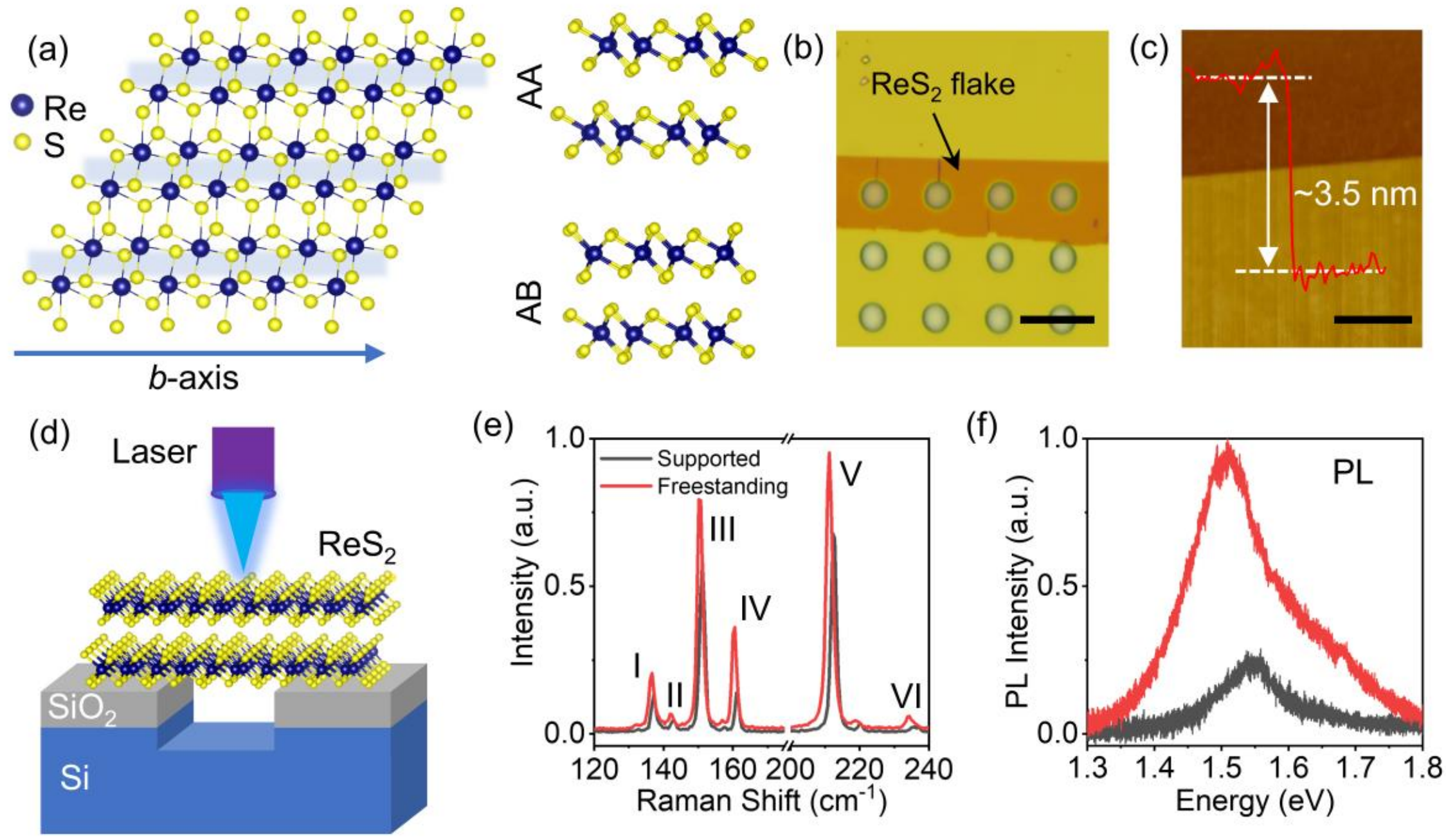

***Figure 1.*** ***(a)*** *Top view of the crystal structure of the $ReS_2$ monolayer. The direction of the Re chains (b-axis) is highlighted. Side views of AA and AB stacked bilayer $ReS_2$ are shown.* ***(b)*** *Optical micrograph of ~3.5 nm $ReS_2$ flake on the array of holes in $SiO_2$/Si substrate. The flake has a sharp edge along the b-axis, which is denoted by a double-headed arrow (black). The scale bar is 10 µm.* ***(c)*** *AFM image of the sample's edge and height profile. The scale bar is 1 µm.* ***(d)*** *Schematic representation of the experimental setup for the optothermal Raman technique.* ***(e)*** *Point Raman and* ***(f)*** *PL spectra from the supported and freestanding regions of the $ReS_2$ flake, respectively.*

$ReS_2$ is an anisotropic material, which indicates that we should expect variation in the Raman response with crystallographic orientation. To determine the orientation, we performed angle-resolved polarized Raman measurements on the $ReS_2$ flakes (with less than ± 2 degree angular uncertainty). The sample was rotated by an angle (θ) between the incident laser polarization and the cleaved edge (indicated by a black double-headed arrow in Figure 2a). The scattered Raman light was collected into the spectrometer without additional polarization optics. Raman spectra for different angles are shown in Figure 2b. As expected, the Raman peak positions of different modes remain unchanged, but the intensity varies with the angle. Recent studies indicate that mode V has maxima along the Re-chain axis (also called the b-axis)[20,35]. Figure 2c shows a polar plot of the Raman intensity of modes III and V of the $ReS_2$ flake (Figure 2a). Modes V and III show maximum Raman intensities at ~5 °and ~35°, respectively, which enables us to identify the *b*-axis of our sample.

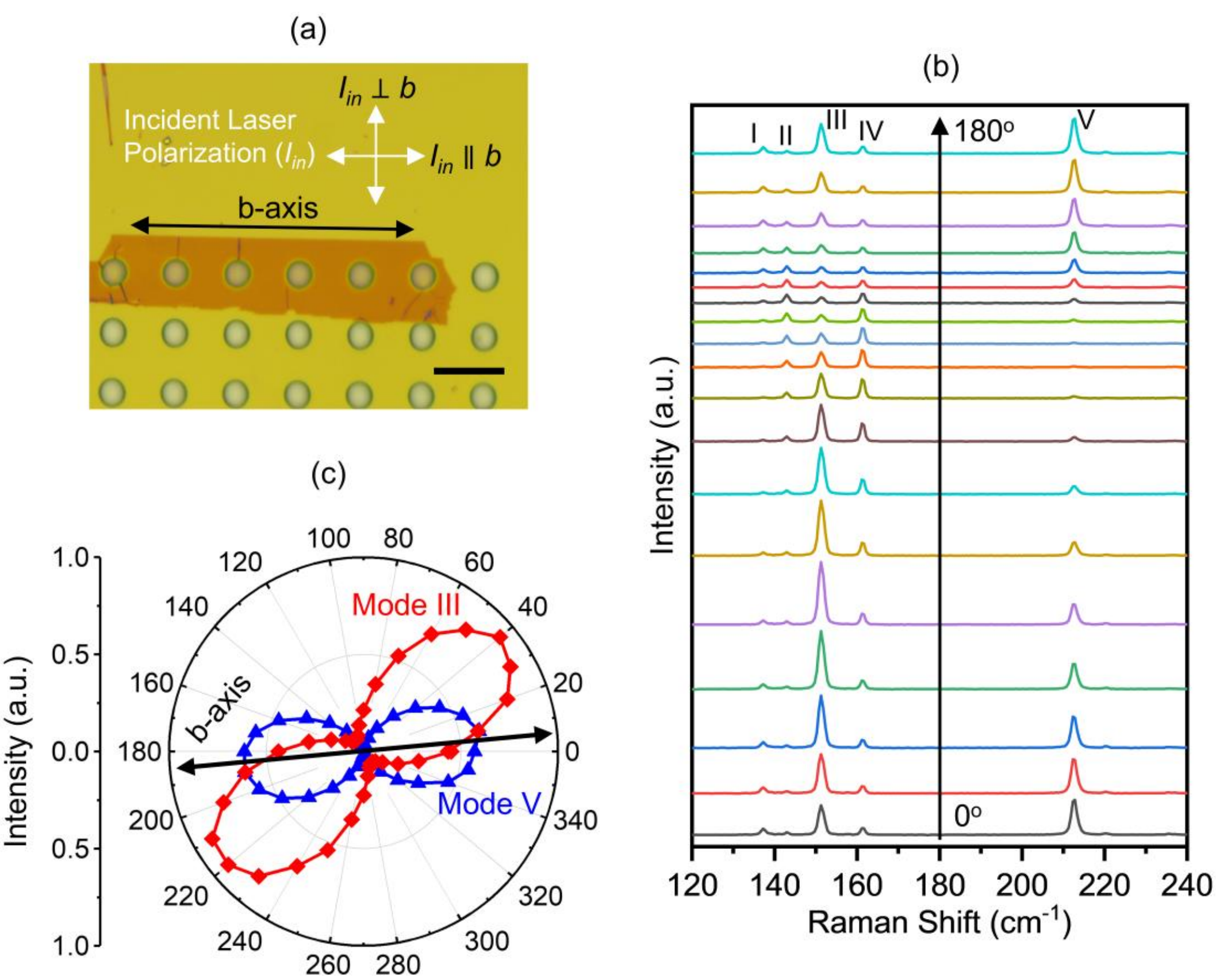

***Figure 2.*** ***(a)*** *Optical micrograph of ~3.5 nm $ReS_2$ flake on the array of holes in $SiO_2$/Si substrate. The flake has a sharp edge along the b-axis, which is indicated by black double arrow. The white double arrow represents the polarization of the 532 nm incident laser. The scale bar is 10 µm.* ***(b)*** *Unpolarized Raman spectra as a function of the sample orientation angle (0° to 180° in steps of 10°).* ***(c)*** *Polar plot of the Raman intensities of modes III (red) and V (blue).*

To measure the thermal conductivity of freestanding $ReS_2$, we used the optothermal Raman technique[36,37]. This technique uses the laser power and temperature-dependent shifts of the Raman modes to estimate the thermal conductivities of freestanding 2D materials. In the case of substrate-supported samples, the substrate underneath can significantly influence the thermal properties of 2D materials. This is mediated by substrate defects, phonon-boundary scatterings, and heat dissipation through the substrate[38,39]. We did not observe any substantial shift in the peak position of Raman spectra on substrate-supported $ReS_2$ in the laser power-dependent measurement owing to faster heat dissipation through $SiO_2$/Si[37] (see Supporting Information S4 for details). The substrate ($SiO_2$/Si) effect can be eliminated using freestanding $ReS_2$ for thermal conductivity measurements. In our measurement, the laser was focused on the center of the freestanding $ReS_2$ flake, and the measurement showed a shift in Raman modes with the change in incident laser power. The Raman mode that is most sensitive to laser power and temperature change is preferred for the anisotropic thermal conductivity measurements. Mode V shows the largest red shifts with laser power and temperature. Therefore, this mode was selected to calculate the anisotropic thermal conductivity of $ReS_2$.

To calculate the thermal conductivity of a 2D flake through the optothermal Raman technique, it is also necessary to know the optical absorbance ($A$) at the specific laser wavelength of the 2D flake. As $ReS_2$ is an anisotropic material, it exhibits large anisotropies in optical absorption values depending on the laser polarization angle with respect to the $b$-axis. For this, we have conducted white light reflection measurements with the incident light polarization parallel and perpendicular to the $b$-axis. More details are in supplementary information S5.

The laser power and temperature-dependent Raman spectra of mode V for five different thicknesses of $ReS_2$ were obtained. The data for ~3.5 nm freestanding $ReS_2$ where incident laser is polarized parallel and perpendicular to the $b$-axis, which are shown in Figure 3a-b and Figure 3c-d, respectively. The Raman mode V of the freestanding $ReS_2$ was redshifted with increasing laser power or temperature. The accurate peak positions of the Raman modes were extracted by Voigt function fitting. In Figure 3e, the power-dependent shift of mode V with incident laser polarization parallel and perpendicular to the $b$-axis is characterized by the expression,

$$\Delta\omega = \omega(P_2) - \omega(P_1) = \gamma \Delta P \quad (1)$$

where $\gamma$ is the slope ($\delta\omega/\delta P$) of the power dependence of the Raman peak position in the linear region and is called the first-order power coefficient. The extracted first-order power coefficient of mode V is found to be -10.89 ± 0.22 $cm^{-1}$/mW, and -7.99 ± 0.24 $cm^{-1}$/mW when laser is polarized parallel and perpendicular to the $b$-axis, respectively. Note that the extracted $\delta\omega/\delta P$ values for incident laser polarization perpendicular and parallel to the $b$-axis exhibit clear differences, arising from the combined effects of anisotropic optical absorption and anisotropic in-plane thermal conductivity.

Similarly, the temperature-dependent frequency shift of mode V with incident laser polarization parallel and perpendicular to the $b$-axis for ~3.5 nm $ReS_2$ shows a linear trend (Figure 3f). The first-order temperature coefficient is extracted using the equation[40,41],

$$\omega(T) = \omega_0 + \chi T \qquad (2)$$

where $\omega_0$ is the frequency of mode V at absolute zero temperature, and $\chi$ is the first-order temperature coefficient of the Raman mode V. For the ~3.5 nm thick flake, the extracted temperature coefficients were $\chi_{||b}$ = -(1.23 ± 0.10) ×10$^{-2}$ cm$^{-1}$/K and $\chi_{\perp b}$ = -(0.69 ± 0.02) ×10$^{-2}$ cm$^{-1}$/K. The temperature coefficient with incident laser polarization parallel to the $b$-axis is observed to have a larger value than perpendicular to the $b$-axis due to anisotropic thermal expansion during the heating process[25].

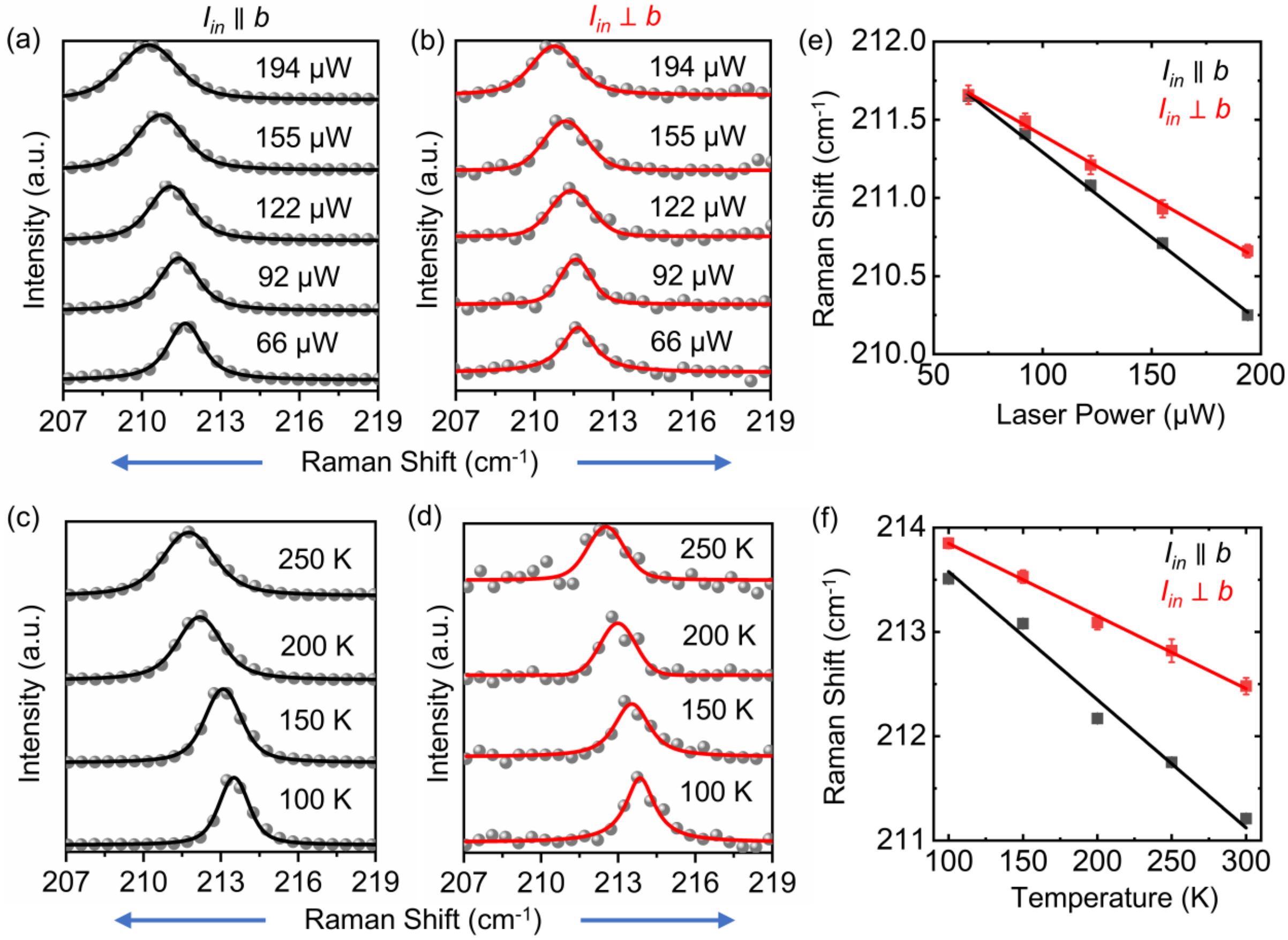

***Figure 3.*** *Power-dependent µ-Raman spectra of mode V for ~3.5 nm thick* $ReS_2$ *(AA) with incident laser polarization* ***(a)*** *parallel and* ***(b)*** *perpendicular to the b-axis. Temperature-dependent micro-Raman spectra of mode V of the same sample with incident laser polarization* ***(c)*** *parallel and* ***(d)*** *perpendicular to the b-axis.* ***(e)*** *Raman peak position vs. laser power plot extracted from the Voigt function fitting of* ***(a)*** *and* ***(b)****. The lines (red and black) are linear fitting to obtain the first-order power coefficient.* ***(f)*** *Raman peak position vs. laser power plot extracted from the Voigt function fitting of* ***(c)*** *and* ***(d)****. The lines (red and black) are linear fitting to obtain the first-order temperature coefficient.*

The expression for thermal conductivity can be derived from the heat conduction equation:

$$\frac{\partial Q}{\partial t} = -\kappa \oint \nabla T . dA \tag{3}$$

$\frac{\partial Q}{\partial t}$ is the thermal power flowing through a cross-sectional area, $\nabla T$ is the temperature gradient, and $\kappa$ is the thermal conductivity of the sample. Assuming radial heat flow from the center to the edge of the freestanding area, the above equation results in the expression[36]:

$$\kappa = \frac{1}{2\pi h}\frac{\Delta P}{\Delta T} \tag{4}$$

where, $h$ is the thickness of the $ReS_2$ flake, and $\Delta T$ is the change in temperature due to the change in heating power $\Delta P$. For anisotropic 2D materials, the local temperature is determined by both the tensor components of thermal conductivity. Therefore, the conductivities along the two directions are extracted after correcting for polarization dependent absorbance. This approach follows previously reported Raman thermometry measurement on anisotropic 2D materials[24,25,42–45]. Using equation 2, the thermal conductivity expression can be modified as,

$$\kappa = A.\chi\left(\frac{1}{2\pi h}\right)\left(\frac{\delta\omega}{\delta P}\right)^{-1} \tag{5}$$

Where $A$ is the absorbance of the $ReS_2$ flake at 532 nm wavelength, $\delta\omega/\delta P$ is the first-order power coefficient, and $\chi$ is the first-order temperature coefficient. Using the values of absorbance, first-order temperature coefficient and power coefficient of the Raman mode V, the thermal conductivities are calculated to be: $\kappa_{||b}$ = 6.4 ± 0.6 W/m-K, and $\kappa_{\perp b}$ = 4.2 ± 0.2 W/m-K for the ~3.5 nm thick $ReS_2$ of AA stacking. The calculated $\kappa_{||b}$ was higher than $\kappa_{\perp b}$, indicating an anisotropic feature in the thermal conductivity. This originates mainly from anisotropic phonon dispersion [25].

To study the influence of stacking order on thermal conductivity, we performed similar measurements on another ~3.5 nm $ReS_2$ flake with AB stacking (see Figure S6a). The *b*-axis of the flake was identified through angle-resolved polarized Raman measurements, following the same procedure used for the AA-stacked sample (see Supporting Information S6 for more details). Laser power and temperature-dependent Raman measurements were then performed with the incident laser polarization oriented parallel and perpendicular to the *b*-axis, and the corresponding peak shifts are shown in Figure 4a-b (see Supporting Information S7 for more details). The Raman peak shifts were analyzed using linear fitting to extract the first-order power and temperature coefficients. For mode V, the first-order power coefficients ($\delta\omega/\delta P$) were obtained as -12.6 ± 0.4 $cm^{-1}$/mW and -8.45 ± 0.19 $cm^{-1}$/mW for laser polarization parallel and perpendicular to the b-axis, respectively. The corresponding temperature coefficients were found to be $\chi_{||b}$ = -(1.17 ± 0.06) ×$10^{-2}$ $cm^{-1}$/K and $\chi_{\perp b}$ = -(0.69 ± 0.03) ×$10^{-2}$. Using these experimentally extracted parameters, the thermal conductivity of the ~3.5 nm AB-stacked $ReS_2$ flake was estimated to be $\kappa_{||b}$ = 5.23 ± 0.37 W/m-K, and $\kappa_{\perp b}$ = 3.98 ± 0.21 W/m-K, confirming pronounced in-plane anisotropic thermal conductivity in the AB-stacked sample as well.

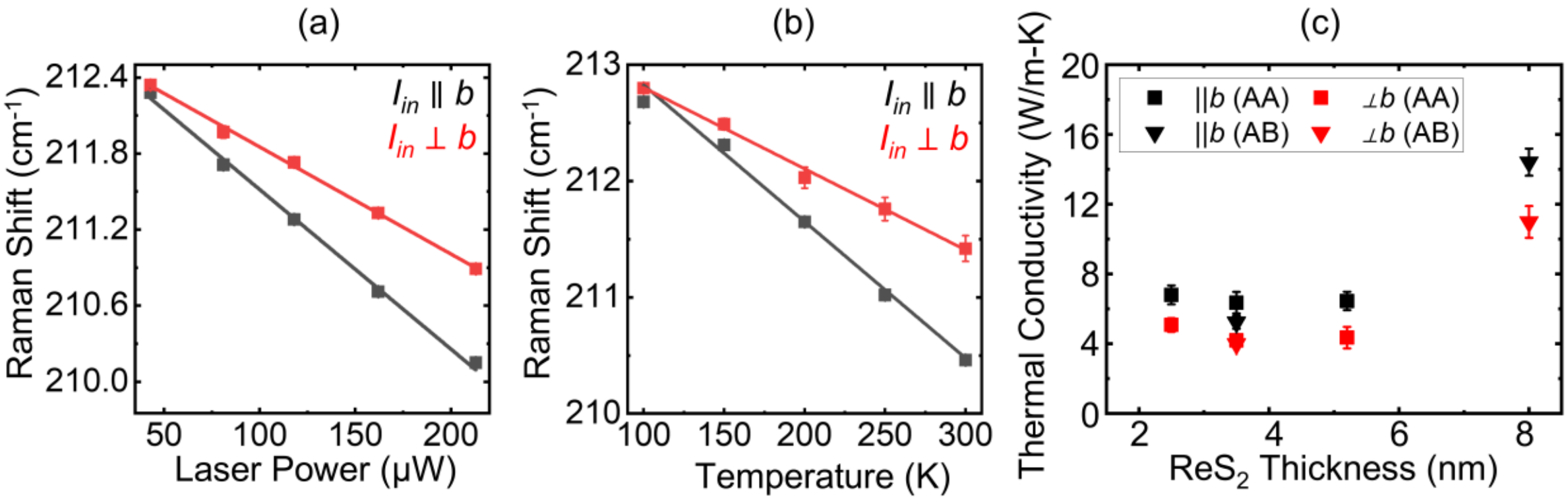

***Figure 4.*** *Raman peak position vs.* ***(a)*** *laser power and* ***(b)*** *temperature plots for mode V of ~3.5 nm* $ReS_2$ *(AB) with incident laser polarization parallel (black) and perpendicular (red) to the b-axis. The*

*lines (red and black) are linear fitting to get the first-order power coefficient.* ***(c)*** *In-plane anisotropic thermal conductivity of $ReS_2$ flakes over a thickness range of 2.5 to 8 nm obtained in this work.*

A direct comparison between the two stacking configurations reveals that, for the same thickness, the AB-stacked $ReS_2$ flake suggests lower in-plane thermal conductivity than the AA-stacked flake along both directions. This observation directly reveals the crucial influence of stacking order and interlayer coupling on phonon transport in low-symmetry layered materials. To the best of our knowledge, this is the first experimental evidence of stacking-order-dependent in-plane thermal conductivity in few-layer $ReS_2$. These findings demonstrate that, beyond the intrinsic crystal anisotropy, subtle variations in atomic stacking can strongly modify phonon dispersion and scattering processes, thereby significantly altering heat transport in few-layer $ReS_2$.

Previous thermal conductivity measurements on 2D materials have indicated a thickness dependence[24,25,46–49]. To understand this effect, we measured thermal conductivity on multiple samples of varying thicknesses (see Supporting Information S8-S10 for more details). We observed a clear nonmonotonic variation in thermal conductivity with thickness (see Figure 4c). This observation is consistent with the thermal conductivities of other 2D materials, including $MoS_2$[46,47], BP[25], and Td-$WTe_2$[24]. This decrease in thermal conductivity with thickness has previously been attributed to changes in the anharmonic force constant and phonon dispersion variations. In $MoS_2$, the anharmonic force constant change is due to interlayer coupling-induced symmetry breaking. However, unlike the $MoS_2$ monolayer, monolayer $ReS_2$ has a distorted 1T phase, which lacks symmetry. Additional weak interlayers are unlikely to impact this force constant. This indicates that the initial reduction in thermal conductivity with increasing thickness of $ReS_2$ is likely due to phonon dispersion changes. Beyond the minima, we also observe an increase in thermal conductivity.

To support our experimental findings, DFT calculations were carried out on trilayer $ReS_2$ with both AA and AB stackings. Although the simulated thickness is at the atomic scale, these calculations provide useful qualitative understanding of the experimentally observed trends. Previous analysis of thermal transport in monolayer and bilayer AA and AB stacked $ReS_2$[50] shows that the AA structure shows higher thermal conductivity than the AB structure for both monolayer and bilayer cases. It was also observed that the bilayer has higher thermal conductivity than the monolayer, indicating that thermal conductivity increases with the number of layers.

In the present work, we further construct a trilayer of optimized $ReS_2$ with AA and AB stackings. Figure S12 (a(b)) and Figure S12 (c(d)) show side (top) views of the AA and AB stackings of optimized tri-layered $ReS_2$, respectively. The optimized in-plane lattice parameters, crystallographic angles, and the inter-layer distances are shown in Table S3. It can be seen that the inter-layer distances change by about ~4.5%, indicating differences in inter-layer coupling between the two structures. Figure 5a-b shows the phonon dispersions of $ReS_2$ with AA and AB stackings, respectively. The positive phonon frequency modes indicate both structures are dynamically stable. It is to be noted that the phonon bands in the AB stacking are slightly flatter than those in the AA stacking. This implies that AA stacking has slightly higher group velocity than AB stacking.

It was experimentally observed that both AA and AB stacking show anisotropic thermal conductivity. To analyze this difference, we examine the phonon group velocity ($v_{g,j\boldsymbol{q}} = \nabla_{\boldsymbol{q}} \omega_{j\boldsymbol{q}}$) in both structures. The phonon group velocity is a harmonic quantity and provides a measure of the movements of phonons inside a given material. Figure 5c-d shows the square phonon group velocities in two orthogonal directions ($\hat{x}$ and $\hat{y}$) as a function of phonon frequencies, for AA and AB stackings, respectively, sampled on a $24 \times 24 \times 1$ mesh. These plots clearly show the differences in the group velocity along two orthogonal directions.

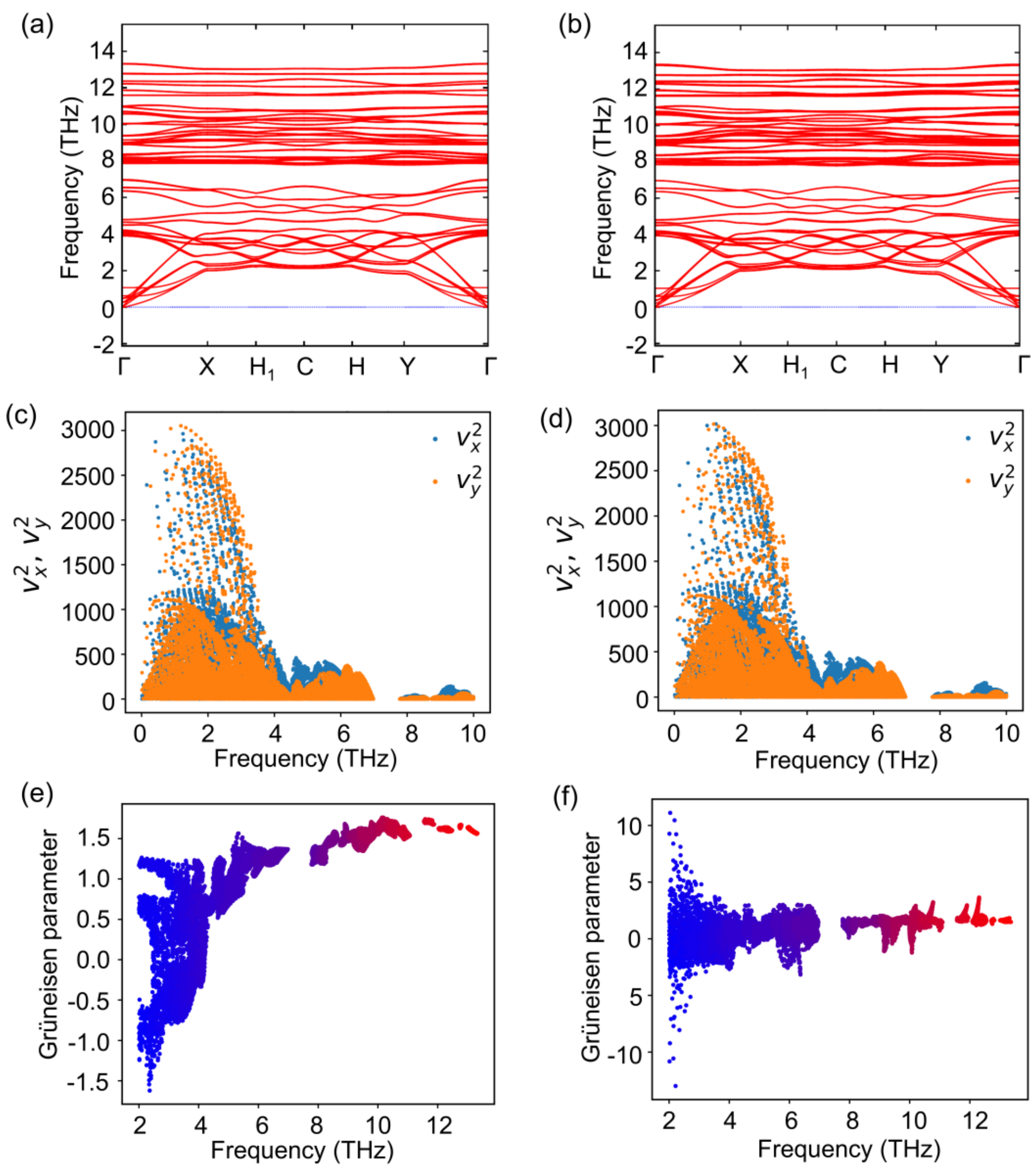

***Figure 5.*** *Phonon dispersions of $ReS_2$ trilayer with* ***(a)*** *AA stacking and* ***(b)*** *AB stacking. The square of phonon group velocity along two orthogonal directions for the* ***(c)*** *AA stacking and* ***(d)*** *AB stacking. Mode Grüneisen parameters for* ***(e)*** *AA stacking and* ***(f)*** *AB stacking. The blue-to-red color gradient in (e, f) indicates frequencies increasing from lower to higher values.*

The linearized phonon Boltzmann Transport Equation (BTE) under the Relaxation Time Approximation (RTA), the lattice thermal conductivity along the direction$(\alpha, \beta)$ is given by expression[51],

$$\kappa_{\alpha\beta} = \frac{1}{NV_0} \sum_{jq} C_{jq}\, \tau_{jq} v_{jq,\alpha} v_{jq,\beta} \delta(\omega' - \omega_\lambda) \tag{6}$$

where $N$ is the number of q-points sampled, $V_0$ is the volume of the primitive cell, $(j,\boldsymbol{q})$ corresponds to a phonon mode, $C_{jq}$ is the modal heat capacity, $\tau_{jq}$ is the phonon lifetime and $v_{jq,\alpha(\beta)}$ is the group velocity in the direction $\alpha(\beta)$.

From equation 6, it can be clearly seen that, along a given direction $\hat{\imath}$, the thermal conductivity tensor component ($\kappa_{ii}$) has a square dependence on the group velocity, leading to the anisotropy observed in in-plane thermal conductivity.

Further, to understand thermal transport across different stackings, we analyze the mode Grüneisen parameter. Figure 5e-f shows the calculated mode Grüneisen parameters as a function of phonon frequency, sampled on a $24 \times 24 \times 1$ mesh for AA and AB stackings, respectively. It can be observed that the Grüneisen parameters (absolute values) are much higher for AB stacking than for AA stacking. The absolute value of the Grüneisen parameter is a measure of the anharmonicity of the system. As the system becomes more anharmonic, phonon-phonon interactions increase, leading to lower phonon lifetimes for AB stacking. However, we find that modal heat capacity remains invariant both for AA and AB stackings, as shown in Figure S13a-b, respectively. Moreover, equation 6 indicates that thermal conductivity increases (decreases) with increasing (decreasing) phonon lifetimes, which further suggests that the thermal conductivity of the AB-stacking should be lower than that of the AA-stacking, aligning with the experimental results.

**Conclusion:**

In conclusion, we experimentally investigated the anisotropic in-plane thermal conductivity of freestanding few-layer $ReS_2$ using polarization-resolved optothermal Raman method. Pronounced in-plane thermal anisotropy was observed along and perpendicular to the crystallographic *b*-axis, with a ~3.5 nm AA-stacked $ReS_2$ flake showing thermal conductivity values of 6.4 ± 0.6 and 4.2 ± 0.2 W/m-K, respectively. We further demonstrate that thermal transport in $ReS_2$ depends strongly on both stacking order and thickness. For the same thickness, AA-stacked $ReS_2$ indicates higher thermal conductivity than AB-stacked $ReS_2$, establishing the important role of stacking configuration and interlayer interactions in phonon transport. In addition, the in-plane thermal conductivity shows an apparent non-monotonic dependence on thickness over the 2.5-8 nm range, indicating significant evolution of phonon dispersion and scattering mechanisms in few-layer $ReS_2$. DFT calculations further support experimental observations by revealing anisotropic phonon transport in both stacking configurations, with AA stacking exhibiting relatively higher phonon group velocity and lower anharmonicity than AB stacking. Our work establishes $ReS_2$ as an important low-symmetry platform for understanding anisotropic heat transport and has potential applications in thermal diodes, anisotropic thermoelectric devices, and other platforms that require directional heat flow.

**Methods:**

We follow the protocol for the dry transfer of 2D materials used by Castellanos-Gomez *et al.*[28]. $ReS_2$ flakes were mechanically exfoliated from bulk crystals (2D semiconductors) using blue tape onto a stamp. The stamp was a thin layer of a commercially available viscoelastic material (Gel-Film from Gel-Pak). Then, the selected flake on the stamp was dry transferred onto the array of holes made in the $SiO_2$/Si substrate using a 2D material transfer system.

AFM (Park NX20) was used to measure the depth of the holes and thicknesses of the $ReS_2$ flakes. Raman and PL measurements were performed using Horiba Scientific Lab RAMHR 800 in backscattering geometry. A 532 nm solid-state diode laser with 0.04 to 0.5 mW power and a ~0.7 μm spot size was focused onto the substrate using a 100× objective. The laser power was chosen as a compromise between a reasonable signal-to-noise ratio (SNR) and laser-induced heating in the $ReS_2$ flakes during temperature-dependent measurements.

**Computational Details:**

All first-principles calculations are performed using density functional theory (DFT) as implemented in the Vienna Ab initio Simulation Package (VASP)[52,53]. The exchange-correlation interactions are described within the generalized gradient approximation (GGA) using the Perdew-Burke-Ernzerhof (PBE) functional[54], and the projector augmented wave (PAW)[55,56] potentials are used to describe electron-ion interactions. Plane wave basis sets with an energy cutoff of 500 eV are used to expand the single-particle eigenstates of the Kohn-Sham equations. Weak vdW interaction is incorporated by Grimme's DFT-D2 method[57,58]. The phonon dispersions of the monolayer were calculated using the PHONONPY package[59], with a $2 \times 2 \times 1$ supercell. The quasi-harmonic approximation is used to calculate the Grüneisen parameter[51], in which first phonopy calculations are performed for the relaxed structure. Further, the calculations are done for structures at slightly higher (+1%) and slightly lower (-1%) volumes. These structures are obtained with cutoffs of $10^{-6} eV$ for the energy threshold, and a force cutoff of $10^{-2} eV/Å$ . The Grüneisen parameter is then sampled over a $24 \times 24 \times 1$ mesh using the phonopy code.

**Supporting Information:**

Fabrication of the hole-array substrate; optical, AFM, and Raman characterization of $ReS_2$ flakes with different thicknesses and stacking configurations; comparison between substrate-supported and freestanding samples; polarization-dependent optical absorption measurements; laser power- and temperature-dependent Raman studies with extracted first-order coefficients and anisotropic thermal conductivities; and DFT analysis of trilayer AA- and AB-stacked $ReS_2$.

**Author information:**

Corresponding Author: *Akshay Naik, anaik@iisc.ac.in

**Data Availability**

All data are available upon reasonable request.

**Acknowledgments**

A.N. acknowledges funding support ANRF through grant CRG/2023/001266 and from the MoE, MeitY, and DST Nano Mission through NNetRA. M.P.S. acknowledges the INSPIRE Fellowship. N.S.K., S.M., and A.K.S. thank the Materials Research Center (MRC), the Solid State Structural and Chemistry

Unit (SSCU), and the Supercomputer Education and Research Center (SERC) at the Indian Institute of Science for providing computational facilities. A.S. acknowledge support from the DST-Nanomisson program of the Department of Science and Technology, Government of India (Grant No. DST/NM/TUE/QM-1/2019), and the DST-INSPIRE Fellowship.

**Author Contributions**

M.P.S. and A.N. developed the experimental framework. M.P.S. performed the fabrication and experiments. A.J. performed absorption measurement with M.P.S. with guidance from A.S. N.S.K. and S.M. performed DFT calculation with guidance from A.K.S. M.P.S. performed the data analysis, discussed and prepared the manuscript, with contributions from all authors.

## Supporting Information

# Anisotropic In-plane Thermal Conductivity of Freestanding Few-layer $ReS_2$

Manavendra Pratap Singh[1], Nihal S. Kumar[2], Subhendu Mishra[2], Abhishek Jangid[3], Akshay Singh[3], Abhishek Kumar Singh[2], Akshay Naik[1*]

[1]*Centre for Nano Science and Engineering, Indian Institute of Science, Bengaluru, Karnataka 560012, India*

[2]*Materials Research Centre, Indian Institute of Science, Bengaluru, Karnataka 560012, India*

[3]*Department of Physics, Indian Institute of Science, Bengaluru, Karnataka 560012, India*

**Corresponding author: anaik@iisc.ac.in*

## S1. Fabrication procedure of the hole-array $SiO_2$/Si substrate

We use a highly p-doped Si wafer with ~290 nm thick thermally grown $SiO_2$/Si. We cut a 4-inch wafer into small rectangular pieces and then cleaned these small pieces. We dip these small pieces of $SiO_2$/Si chips into acetone and ultrasonicate it for 3 minutes. Then, these chips are transferred into iso-propyl-alcohol (IPA) and are again sonicated for 3 minutes. We carry out standard optical lithography throughout the work using Mask writer Heidelberg (uPG501). We use AZ5214E positive optical resist. At first, we spin-coated AZ5214E uniformly on the substrate at a rotating speed of 4000 rpm for 40 seconds (including 5 seconds of ramp time). Then, the substrate is heated at 110 °C for 60 seconds on a hot plate to remove the solvent from the resist film. 10×10 arrays of circles (diameter ~4 μm) have been directly written by Mask writer Heidelberg (uPG501) with defocus -2 and exposure time 90 ms. The sample is developed in AZ726MIF developer for 25 seconds and then placed in the DI water, which acts as a stopper. It is then dried out using a $N_2$ blow-drier. The sample was hard-baked for 3 min at 110 °C for dry etch processes. We successfully fabricated the arrays of holes with a depth of ~685 nm and diameter of ~4 μm followed by reactive ion etching (RIE) of $SiO_2$ & Si layers. The atomic force microscopy (AFM) image is attached below in Figure S1.

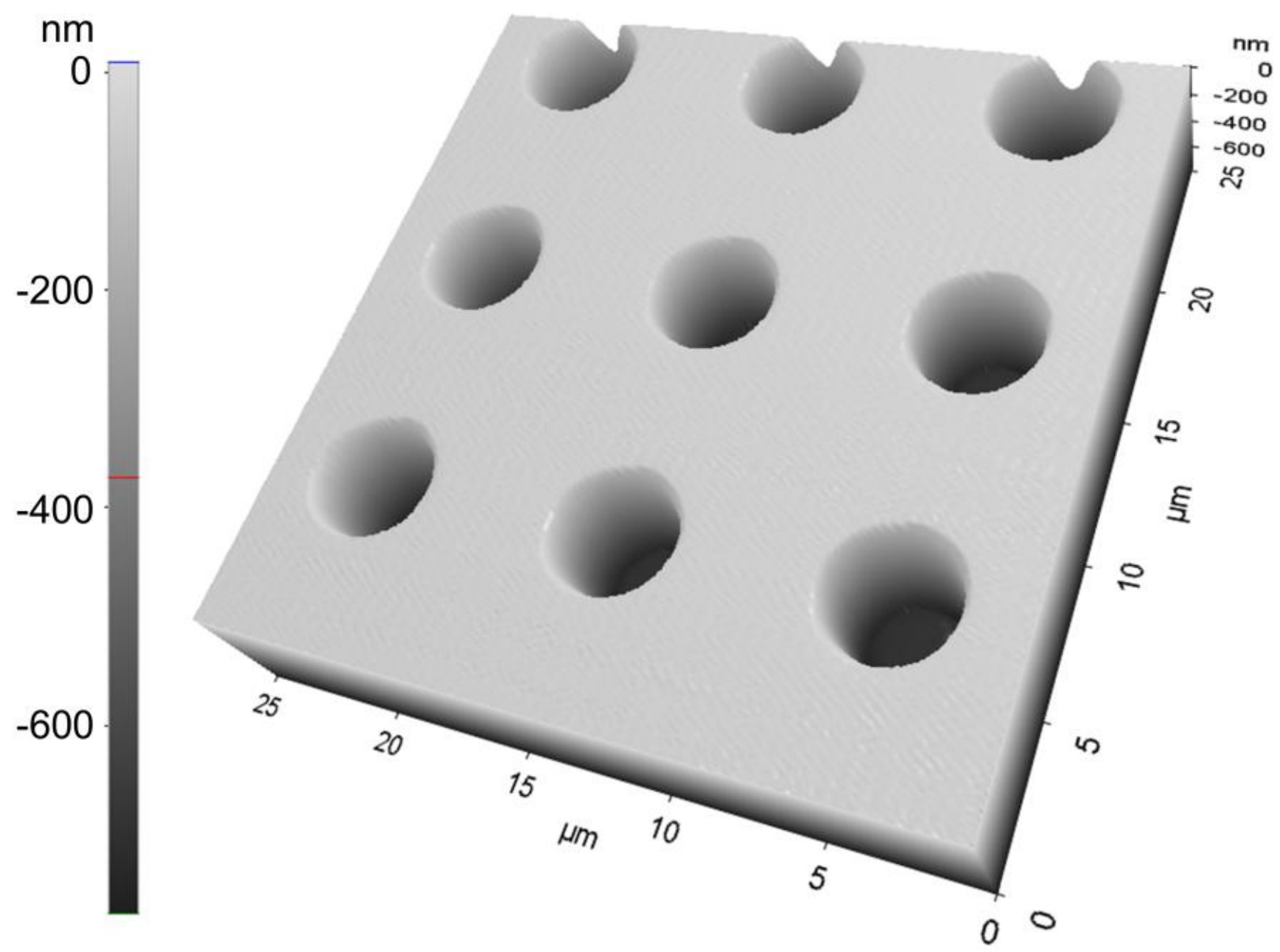

***Figure S1.*** *3D AFM image of the holey substrate.*

**S2. Extended Raman spectra of the $ReS_2$ flake**

Figure S2 shows 18 Raman modes of ~3.5 nm thick $ReS_2$, which were previously reported for $ReS_2$[1].

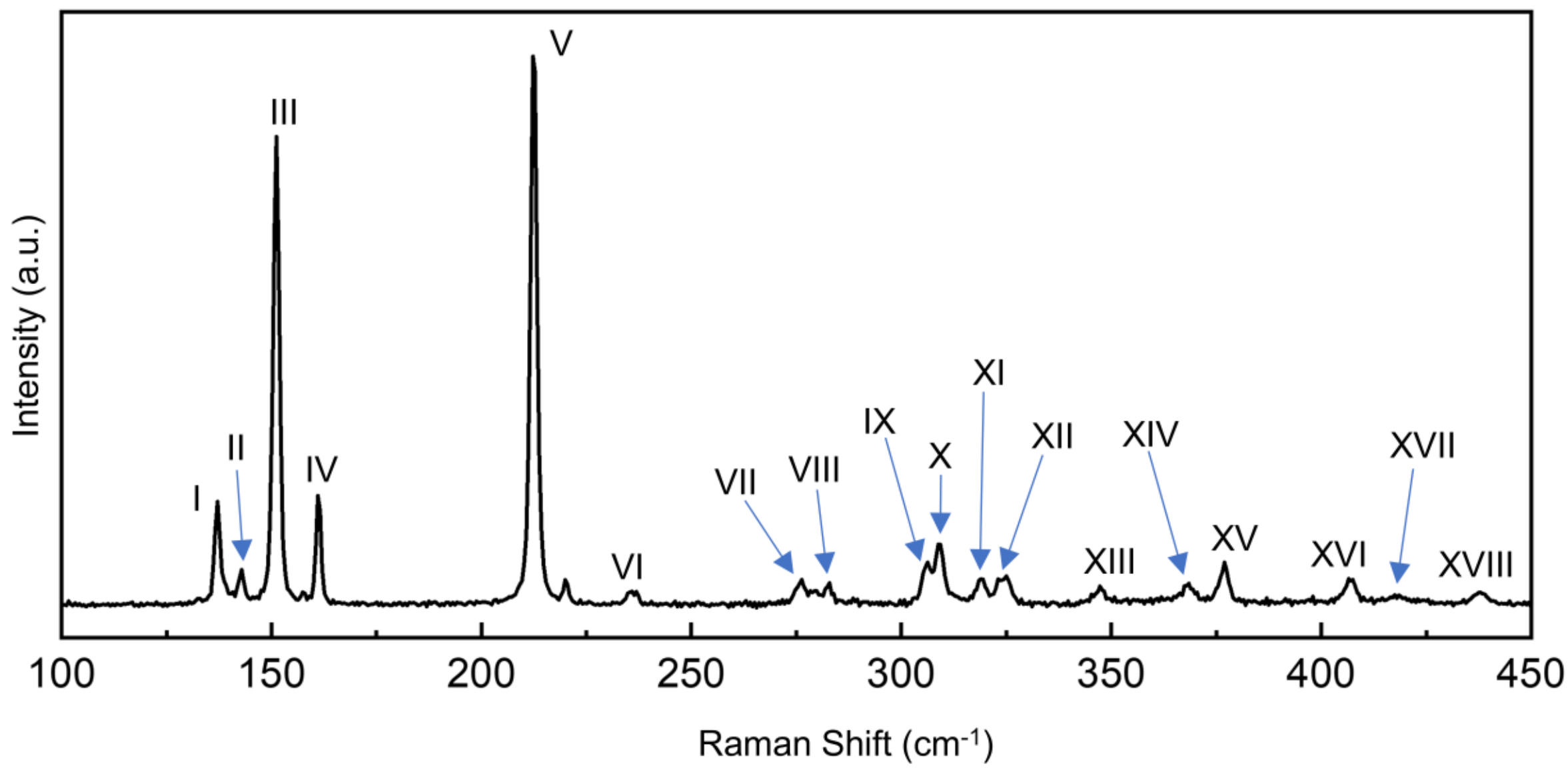

***Figure S2.*** *Raman spectra of ~3.5 nm thick $ReS_2$.*

### S3. Raman confirmation of AA stacking in the ~3.5 nm $ReS_2$ sample

We use the separation between Raman modes I and III to identify whether the stacking order is AA or AB [2–4]. For monolayer $ReS_2$, the peak separation was ~ 17 $cm^{-1}$. The separation was less than that for AA stacking, and for AB stacking, it was more[2,3]. Figure S3 shows that the flake's separation between Raman modes III and I is 14.02 $cm^{-1}$, which confirms the AA stacking order of ~3.5 nm thick $ReS_2$.

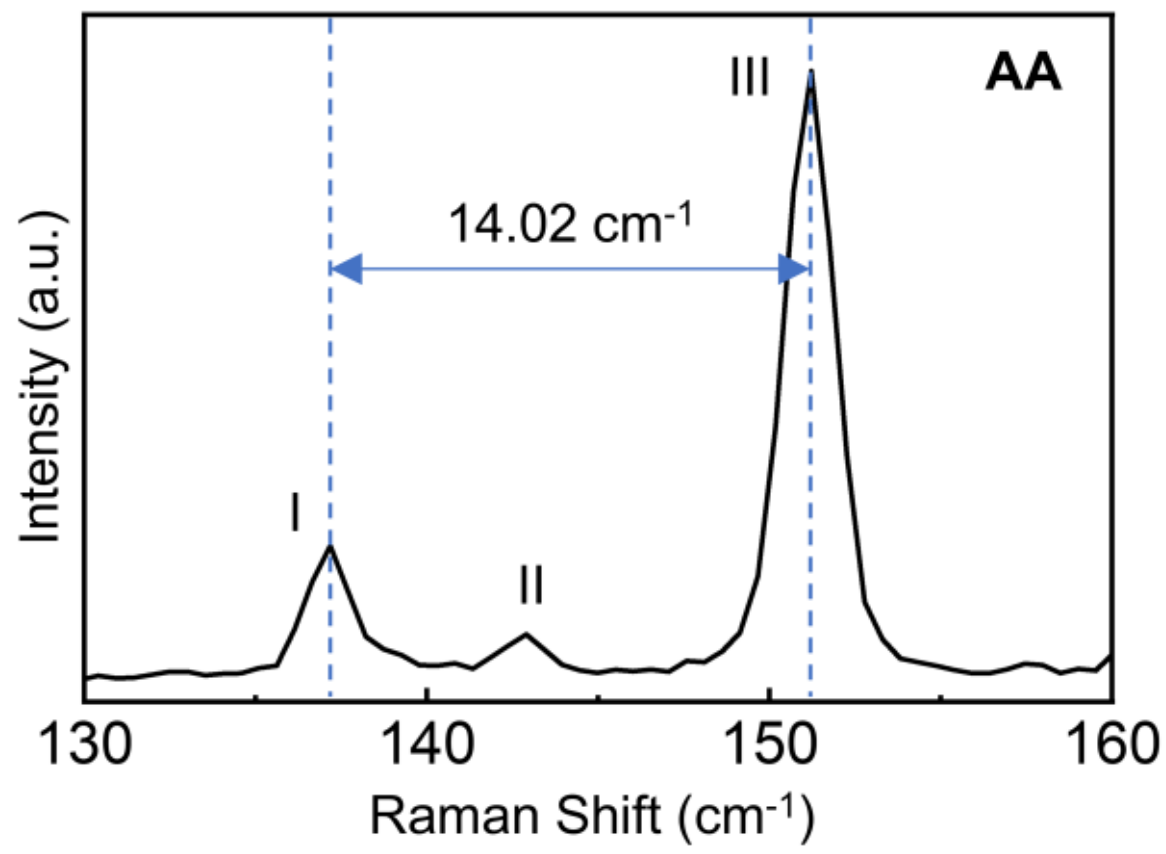

***Figure S3.*** *Raman spectra of AA stacking $ReS_2$ with thickness ~3.5 nm.*

### S4. Comparison of power-dependent Raman spectra from substrate-supported and freestanding $ReS_2$

Figures S4a and S4b show the power-dependent Raman spectra on the substrate-supported and freestanding $ReS_2$. There is no substantial shift in the peak position of Raman modes on substrate-supported $ReS_2$ with laser power variation owing to faster heat dissipation through $SiO_2$/Si[5]. However, the freestanding $ReS_2$ shows a significant red shift with increasing laser power, which is used in calculating thermal conductivity.

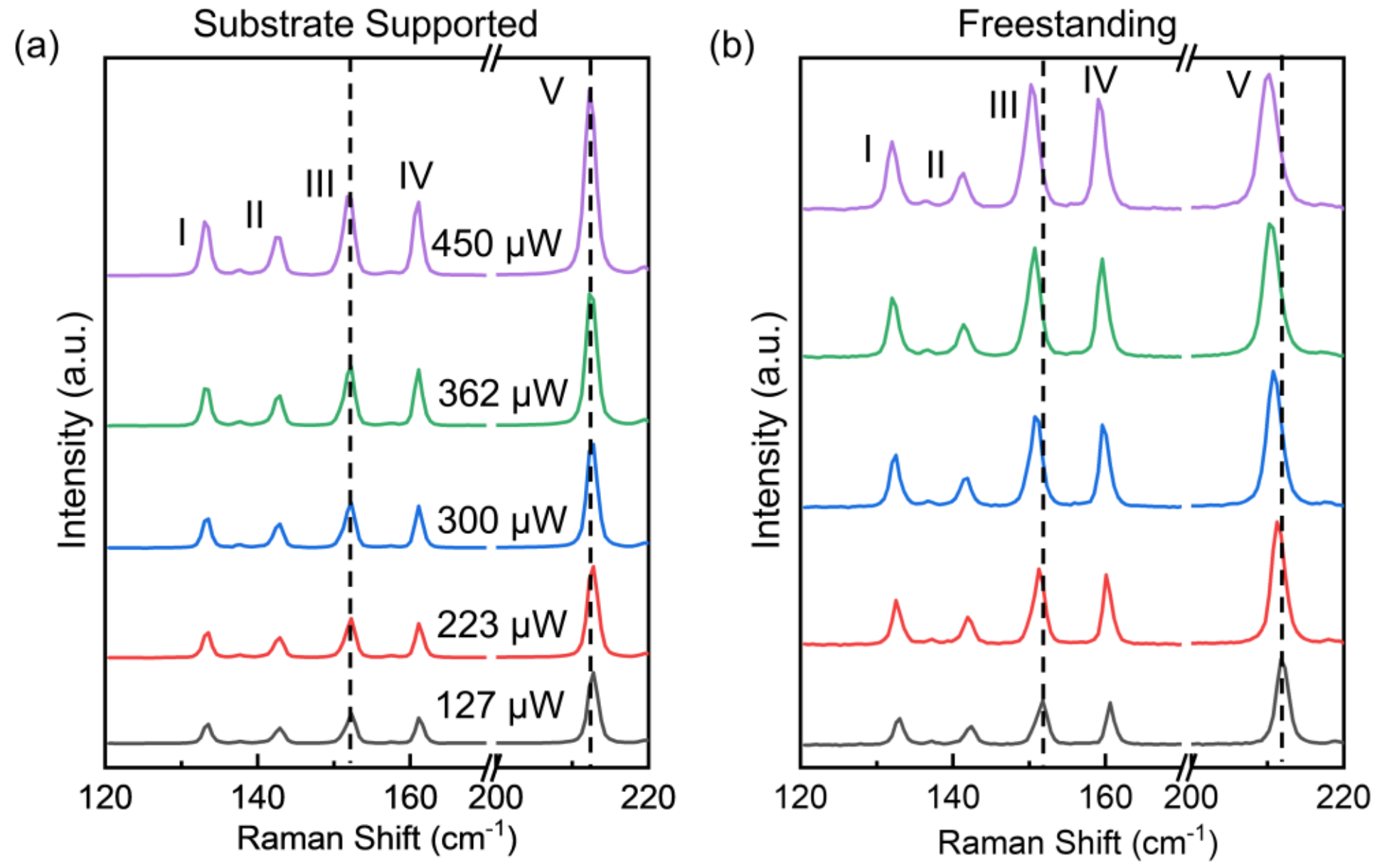

***Figure S4.*** *Power-dependent Raman spectra on* ***(a)*** *substrate supported and* ***(b)*** *freestanding* $ReS_2$*, respectively.*

## S5. Polarization-dependent optical absorption of $ReS_2$

Since $ReS_2$ is an anisotropic material, its optical absorption depends strongly on the polarization of the incident light with respect to the crystallographic *b*-axis. For an accurate estimation of thermal conductivity using the optothermal Raman method, it is therefore necessary to determine the optical absorbance ($A$) of the flake at the excitation wavelength.

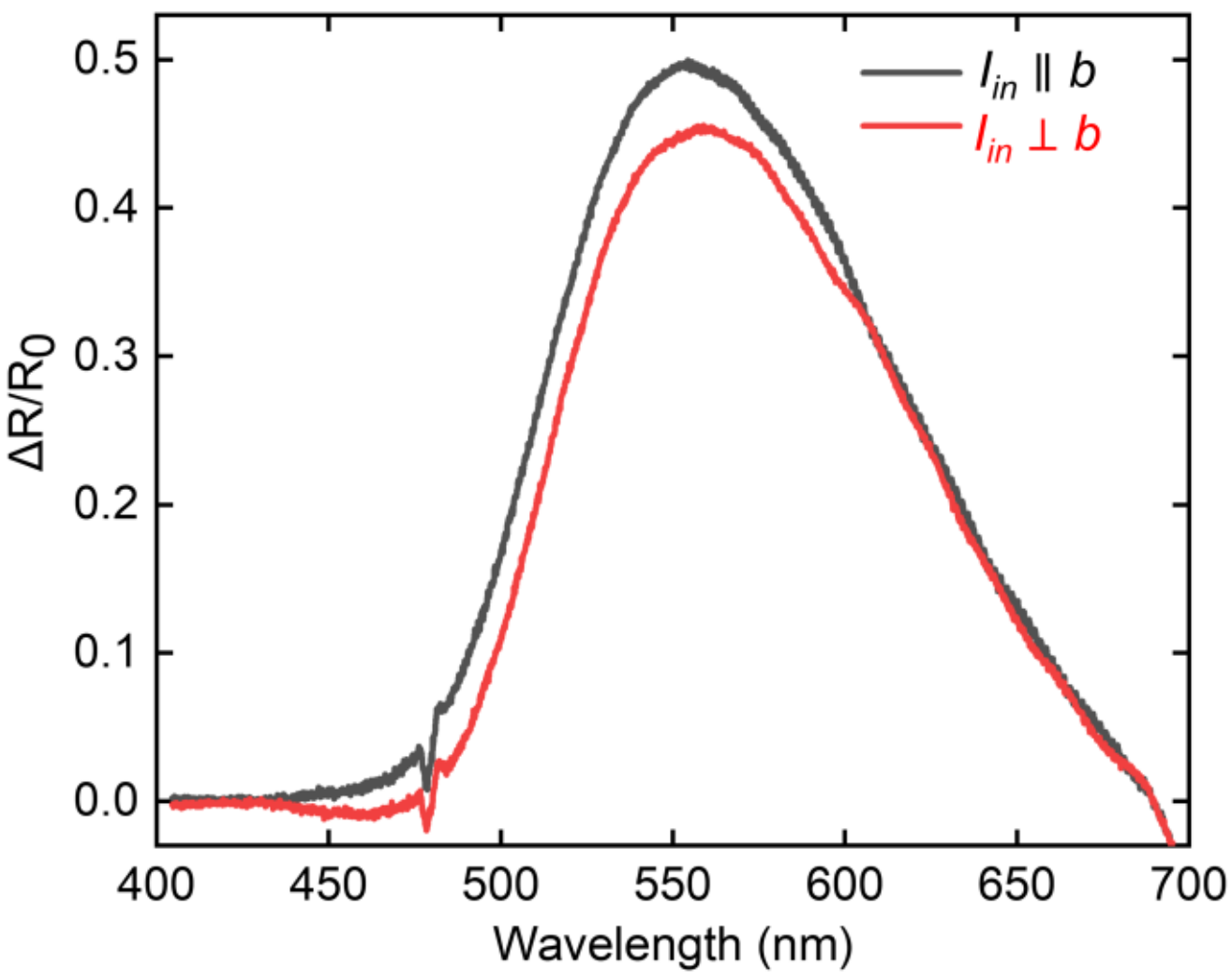

***Figure S5.*** *Reflection contrast when white light polarization is (a) parallel and (b) perpendicular to the b-axis of ~3.5 nm $ReS_2$ flake.*

To account for this anisotropy, white-light reflection measurements were carried out with the incident polarization oriented both parallel and perpendicular to the *b*-axis. As shown in Figure 4.10, at 532 nm, the reflection contrast (ΔR/R0) is 0.44 for incident polarization parallel to the *b*-axis and 0.38 for polarization perpendicular to the *b*-axis.

The optical absorbance, *A,* is calculated using the relation[6],

$$A = \frac{\Delta R}{R_0} . \frac{n_s^2 - 1}{4} \tag{4.1}$$

Where $n_s$ is the refractive index of the substrate. Using $n_s = 1.4607$ at 532 nm, the absorbance is obtained as 12.46% for incident white light polarization parallel to the *b*-axis and 10.76% for the perpendicular direction for the ~3.5 nm thick $ReS_2$ flake.

The absorption coefficient $\alpha$ is then estimated using the relation,

$$A = 1 - e^{-\alpha d} \tag{4.2}$$

Where $d$ is the thickness of the flake. From this, the values of $\alpha$ are calculated to be $\alpha_{\parallel b} = 3.8 \times 10^7 m^{-1}$ and $\alpha_{\perp b} = 3.2 \times 10^7 m^{-1}$at 532 nm.

Using Equation A.8, the optical absorbance values for $ReS_2$ flakes of different thicknesses used in this study are further calculated and summarized in Table S1.

| **Thickness (*d*), nm** | $A_{\parallel b}$ **(%)** | $A_{\perp b}$ **(%)** |
|---|---|---|
| 2.5 | 9.07 | 7.8 |
| 3.5 | 12.46 | 10.76 |
| 5.2 | 17.92 | 15.54 |
| 8 | 26.2 | 22.88 |

***Table S1.*** *Optical absorbance of* $ReS_2$ *flakes of different thicknesses for incident white light polarization parallel and perpendicular to the b-axis.*

**S6. Optical microscopy, Raman, and AFM characterization of the ~3.5 nm AB-stacked $ReS_2$ sample**

Figure S6a shows that the separation between Raman modes III and I of the $ReS_2$ flake is 18.52 $cm^{-1}$, which confirms the AB stacking order of $ReS_2$. Figure S6b shows an optical image of the $ReS_2$ flake on the "holey" substrate of $SiO_2$/Si. The thickness (~3.5 nm) of the transferred flake was measured using atomic force microscopy (AFM) at the edge of the flake (Figure S6c). To find the *b*-axis, we performed angle-resolved polarized Raman measurements of the $ReS_2$ flake, similar to the previous sample. Figure S6d shows a polar plot of the Raman intensity of modes III and V of the $ReS_2$ flake (Figure S6d).

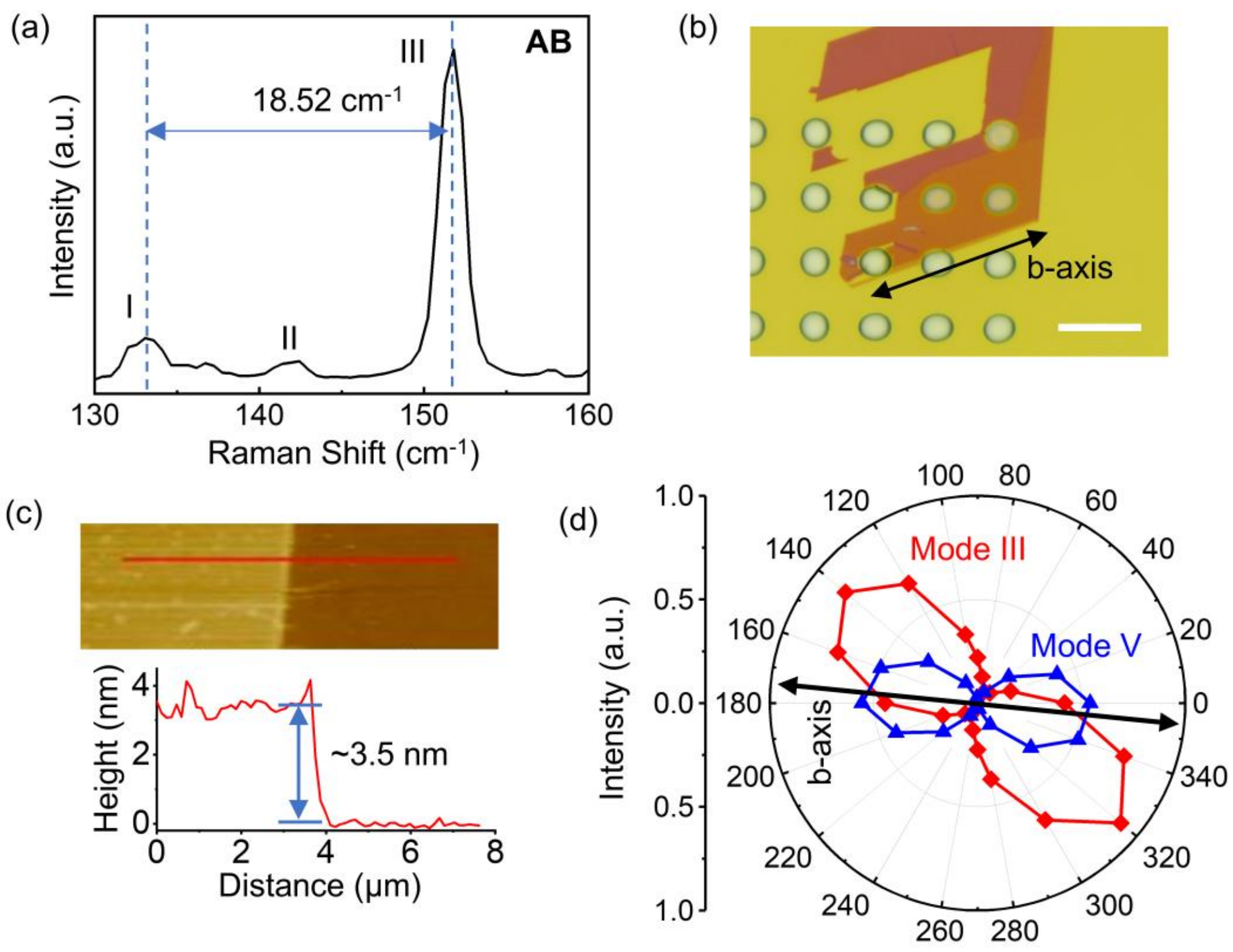

***Figure S6. (a)*** *Raman spectra of AB stacking $ReS_2$ with thickness ~3.5 nm.* ***(b)*** *Optical micrograph of the $ReS_2$ flake on the array of holes in $SiO_2$/Si substrate. The flake has a sharp edge along the b-axis, which is denoted by a double-headed arrow (black). The scale bar is 10 μm.* ***(c)*** *AFM image of the sample's edge and height profile.* ***(d)*** *Polar plot of the Raman intensities of modes III (red) and V (blue).*

## S7. Laser power- and temperature-dependent Raman spectra of mode V for the ~3.5 nm AB-stacked $ReS_2$ sample

The laser power and temperature-dependent Raman spectra of mode V for the ~3.5 nm freestanding $ReS_2$ (AB stacking) along the *b*-axis and cross *b*-axis are shown in Figure S7a-b and Figure S7c-d, respectively. All Raman modes of the freestanding $ReS_2$ were redshifted with increasing laser power or temperature. The accurate peak positions of the Raman modes were extracted by Voigt function fitting.

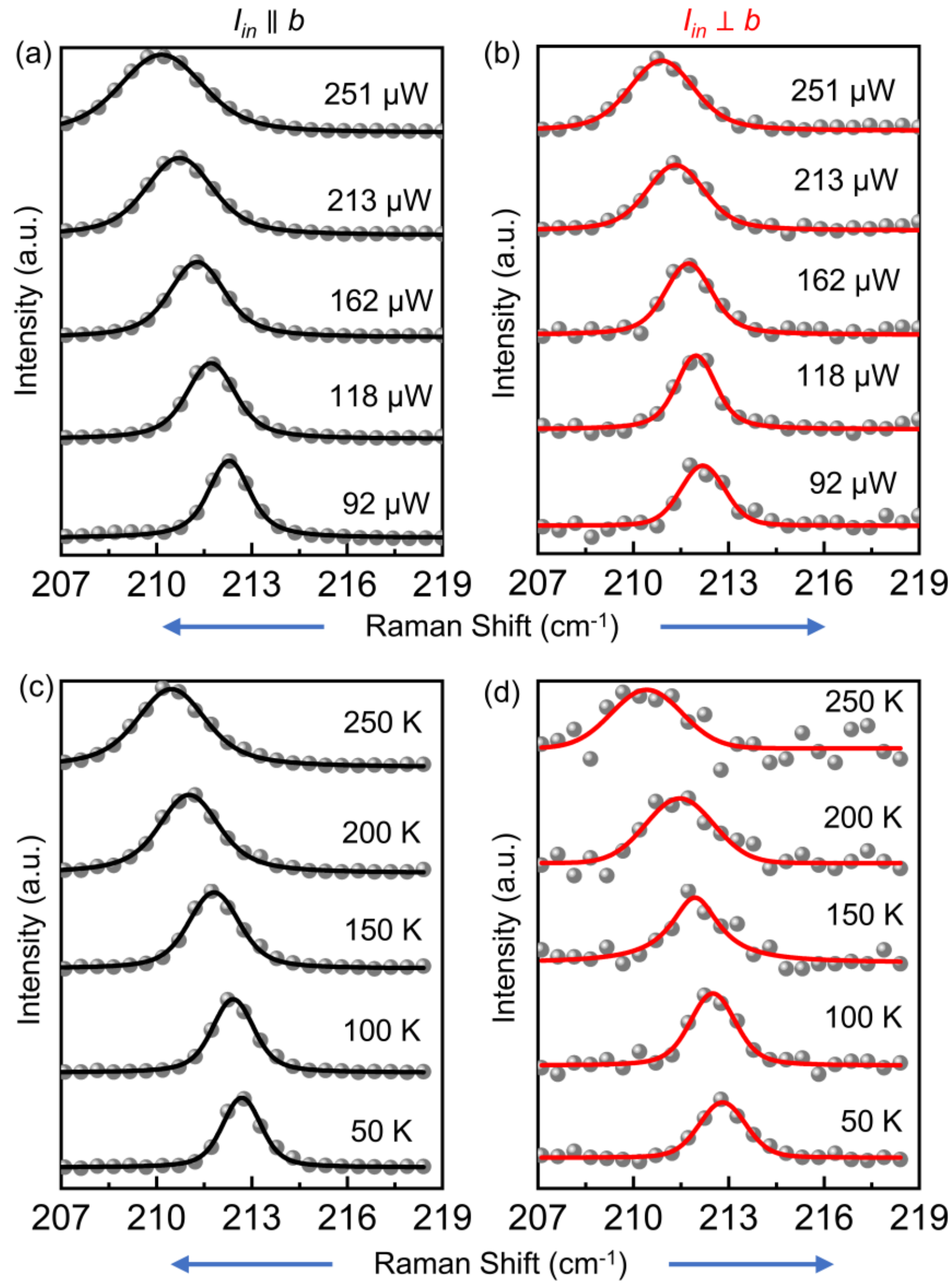

***Figure S7.*** *Power-dependent μ-Raman spectra of mode V for ~3.5 nm thick* $ReS_2$ *(AB) along the* ***(a)*** *b-axis and* ***(b)*** *cross b-axis. Temperature-dependent micro-Raman spectra of mode V of the same sample along* ***(c)*** *b-axis and* ***(d)*** *cross b-axis. All spectra were fitted with the Voigt function.*

## S8. Optical, AFM, and Raman characterization of ~2.5, ~5.2, and ~8 nm $ReS_2$ samples

Figure S8a-c shows optical micrographs of three different $ReS_2$ flakes on the array of holes in the $SiO_2$/Si substrate. The b-axis is denoted by a double-headed arrow (black) on each optical image. Figure S8d-f shows their AFM images of the sample's edges and height profiles. The measured thicknesses are ~2.5 nm, ~5.2 nm, and ~8 nm. Figure S8g-i shows the separation between Raman modes III and I of the $ReS_2$ flakes. For ~2.5 nm thick flake, the separation is 14.89 $cm^{-1}$, which confirms the AA stacking. The flake with thickness ~5.2 nm shows the separation:13.26 $cm^{-1}$, which confirms the AA stacking. For ~8 nm thick flake, the separation is 18.74 $cm^{-1}$, which confirms the AB stacking.

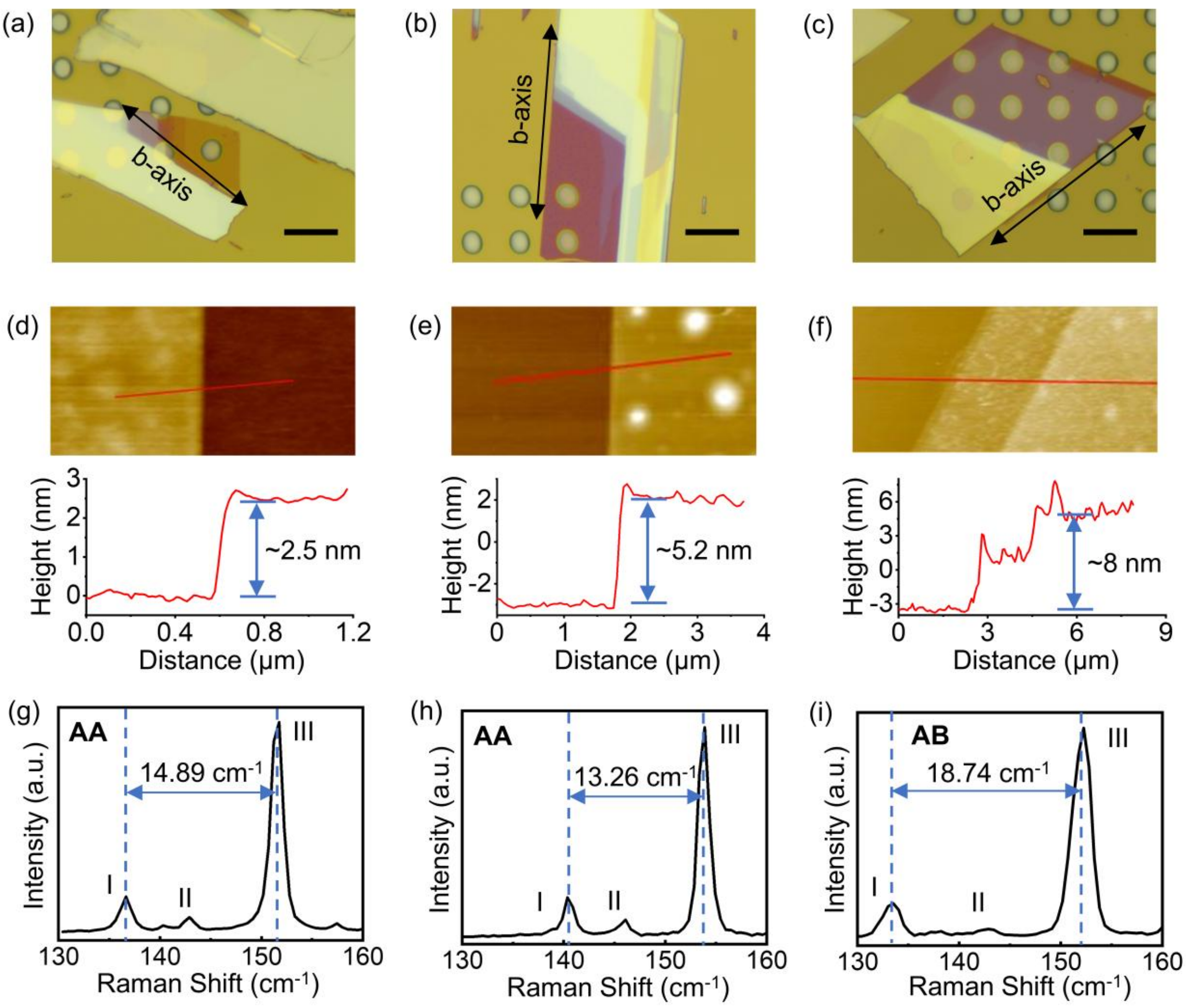

***Figure S8. (a-c)*** *Optical micrograph of the $ReS_2$ flake on the array of holes in $SiO_2$/Si substrate. The b-axis is denoted by a double-headed arrow (black). The scale bar is 10 µm.* ***(d-f)*** *AFM image of the*

*sample's edge and height profile.* ***(g-i)*** *Raman spectra of $ReS_2$ with thicknesses ~2.5 nm (AA stacking), ~5.2 nm (AA stacking), and ~8 nm (AB stacking), respectively.*

## S9. Laser power- and temperature-dependent Raman spectra of mode V for ~2.5 nm (AA), ~5.2 nm (AA), and ~8 nm (AB) $ReS_2$ samples

### (A) Mode V of ~2.5 nm $ReS_2$ (AA)

The extracted first-order power coefficient, temperature coefficient of mode V along the *b*-axis and cross *b*-axis, and the calculated anisotropic thermal conductivities are listed in Table S2.

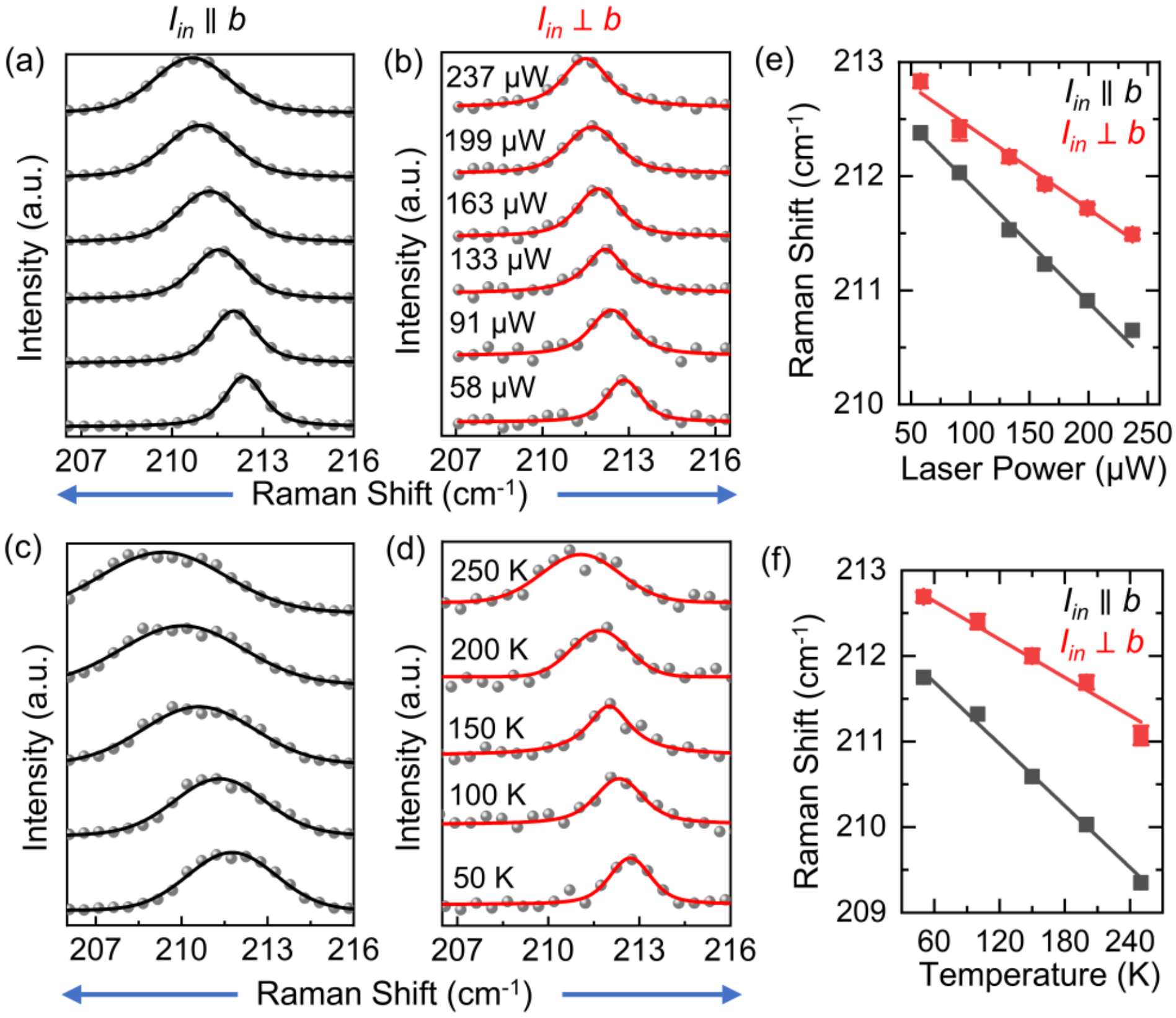

***Figure S9. (a,b)*** *Power-dependent and* ***(c,d)*** *temperature-dependent Raman spectra of mode V for a ~2.5 nm AA-stacked $ReS_2$ flake, measured with incident laser polarization parallel and perpendicular to the b-axis.* ***(e,f)*** *Corresponding Raman peak position as a function of laser power and temperature, respectively, extracted using Voigt fitting; solid lines represent linear fits used to obtain the first-order power and temperature coefficients.*

### (B) Mode V of ~5.2 nm $ReS_2$ (AA)

The extracted first-order power coefficient, temperature coefficient of mode V along the *b*-axis and cross *b*-axis, and the calculated anisotropic thermal conductivities are listed in Table S2.

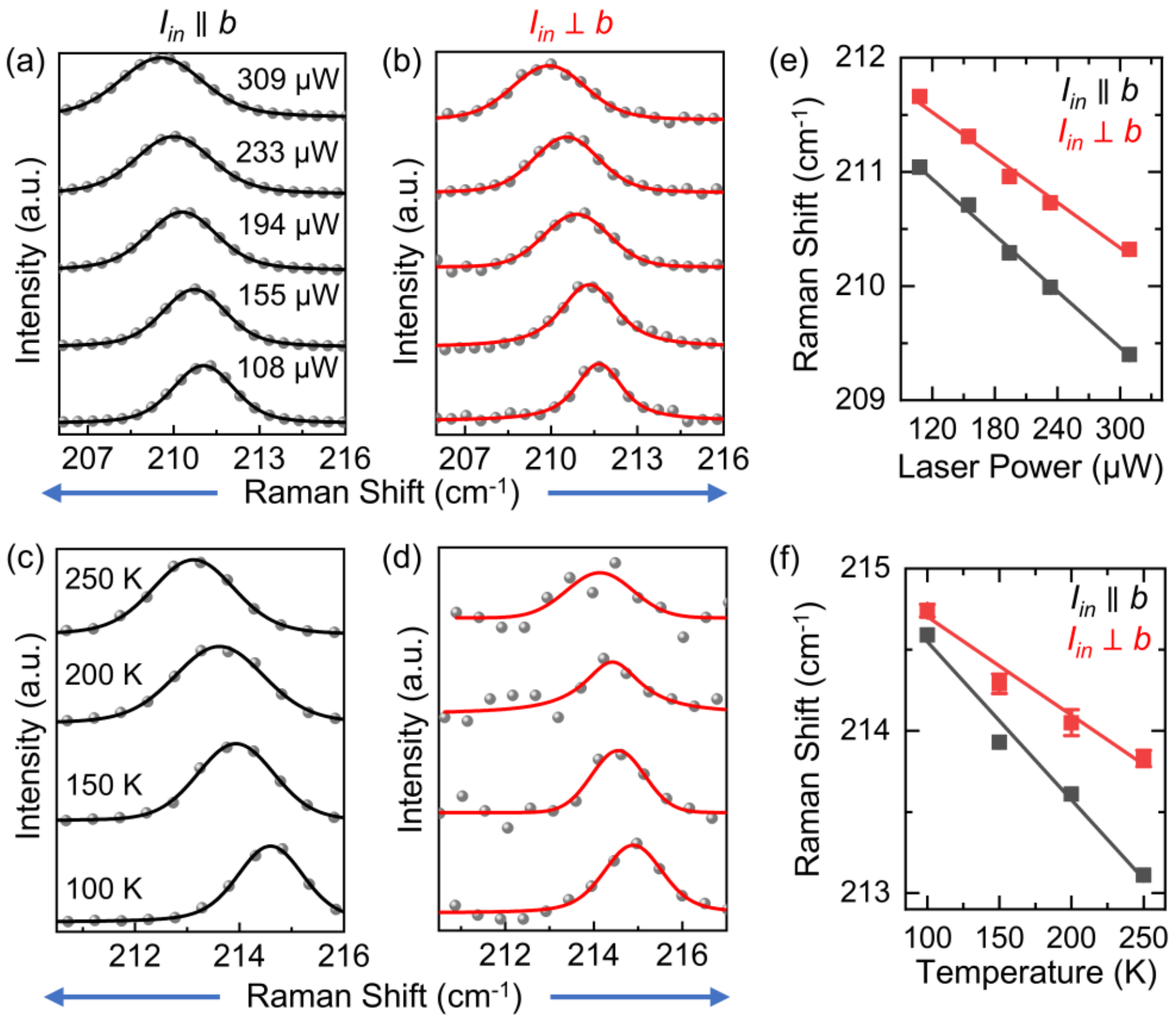

***Figure S10. (a,b)*** *Power-dependent and* ***(c,d)*** *temperature-dependent Raman spectra of mode V for a ~5.2 nm AA-stacked* $ReS_2$ *flake, measured with incident laser polarization parallel and perpendicular to the b-axis.* ***(e,f)*** *Corresponding Raman peak position as a function of laser power and temperature, respectively, extracted using Voigt fitting; solid lines represent linear fits used to obtain the first-order power and temperature coefficients.*

**(C) Mode V of ~8 nm** $ReS_2$ **(AB)**

The extracted first-order power coefficient, temperature coefficient of mode V along the *b*-axis and cross *b*-axis, and the calculated anisotropic thermal conductivities are listed in Table S2.

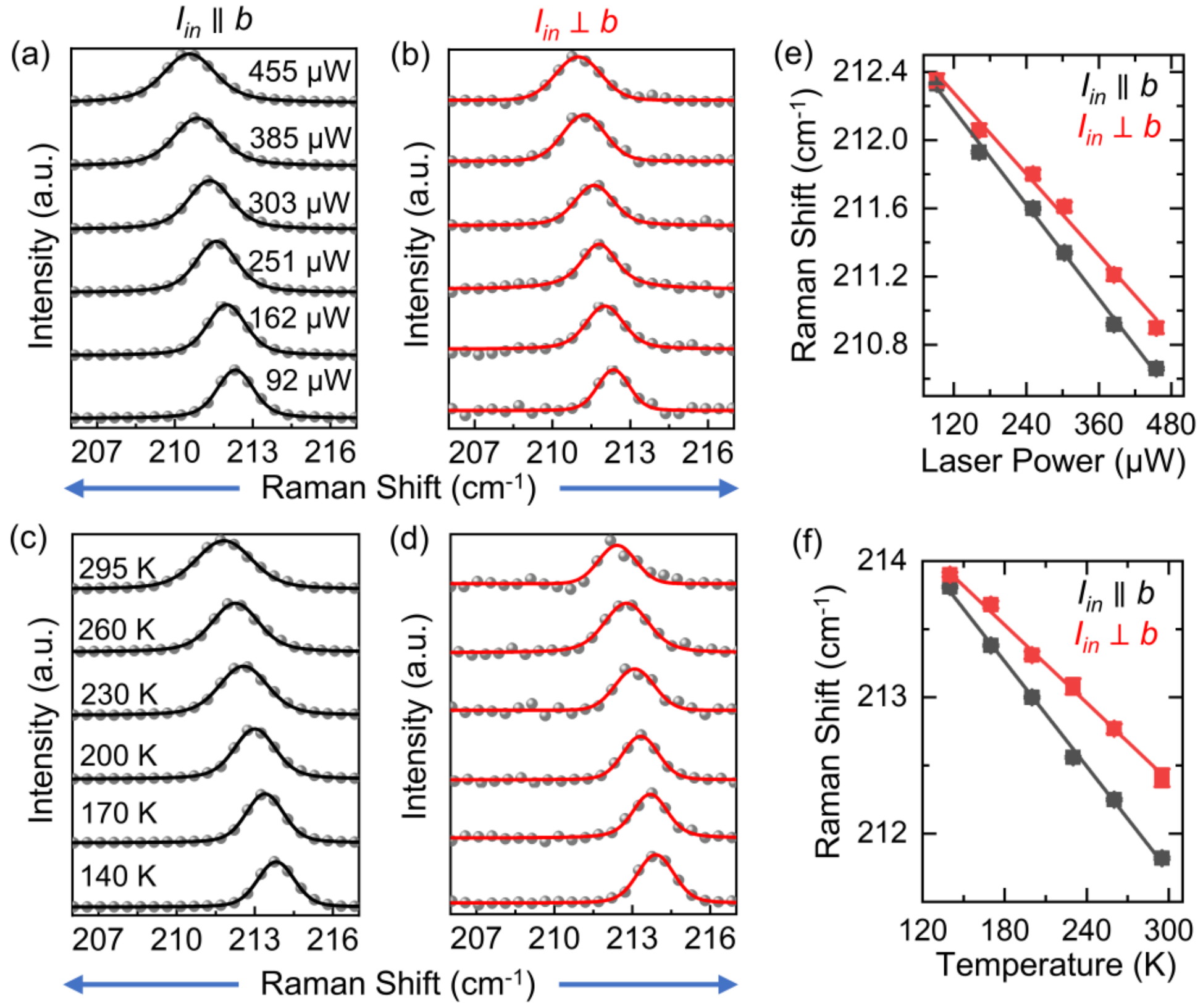

***Figure S11. (a,b)*** *Power-dependent and* ***(c,d)*** *temperature-dependent Raman spectra of mode V for a ~8 nm AB-stacked* $ReS_2$ *flake, measured with incident laser polarization parallel and perpendicular to the b-axis.* ***(e,f)*** *Corresponding Raman peak position as a function of laser power and temperature, respectively, extracted using Voigt fitting; solid lines represent linear fits used to obtain the first-order power and temperature coefficients.*

**S10. First-order power coefficients, temperature coefficients, and anisotropic in-plane thermal conductivities extracted using mode V for all $ReS_2$ samples**

| $ReS_2$ samples thickness and their orientation | | Power Coefficient ($cm^{-1}/mW$) | Temperature coefficient ($cm^{-1}/K$) | Thermal Conductivity (W/m-K) |
|---|---|---|---|---|
| 2.5 nm [AA] | $\|\|b$ | -(10.36 ± 0.36) | -(1.22 ± 0.06) ×$10^{-2}$ | 6.80 ± 0.54 |
| | $\perp b$ | -(7.17 ± 0.45) | -(0.74 ± 0.05) ×$10^{-2}$ | 5.07 ± 0.39 |
| 3.5 nm [AA] | $\|\|b$ | -(10.89 ± 0.22) | -(1.23 ± 0.10) ×$10^{-2}$ | 6.35 ± 0.62 |
| | $\perp b$ | -(7.99 ± 0.24) | -(0.69 ± 0.02) ×$10^{-2}$ | 4.19 ± 0.21 |
| 3.5 nm [AB] | $\|\|b$ | -(12.6 ± 0.40) | -(1.17 ± 0.06) ×$10^{-2}$ | 5.23 ± 0.37 |
| | $\perp b$ | -(8.45 ± 0.19) | -(0.69 ± 0.03) ×$10^{-2}$ | 3.98 ± 0.21 |
| 5.2 nm [AA] | $\|\|b$ | -(8.2 ± 0.26) | -(0.97 ± 0.07) ×$10^{-2}$ | 6.45 ± 0.53 |
| | $\perp b$ | -(6.59 ± 0.41) | -(0.60 ± 0.06) ×$10^{-2}$ | 4.35 ± 0.62 |
| 8 nm [AB] | $\|\|b$ | -(4.61 ± 0.12) | -(1.28 ± 0.03) ×$10^{-2}$ | 14.41 ± 0.78 |
| | $\perp b$ | -(3.96 ± 0.20) | -(0.95 ± 0.03) ×$10^{-2}$ | 10.98 ± 0.91 |

***Table S2.*** *First-order power and temperature coefficients using mode V, and anisotropic thermal conductivities of $ReS_2$ flakes of different thicknesses.*

## S11. DFT analysis of trilayer AA- and AB-stacked $ReS_2$ structures and phonon transport properties

### Structure optimization

Phonon calculations require structures optimized under strict conditions. The optimized structures are obtained by setting an energy threshold of $10^{-6} eV$ and a force cutoff of $10^{-2} eV/$ Å. Figure S12 shows the top and side views of the optimized structures for the AA and AB stackings.

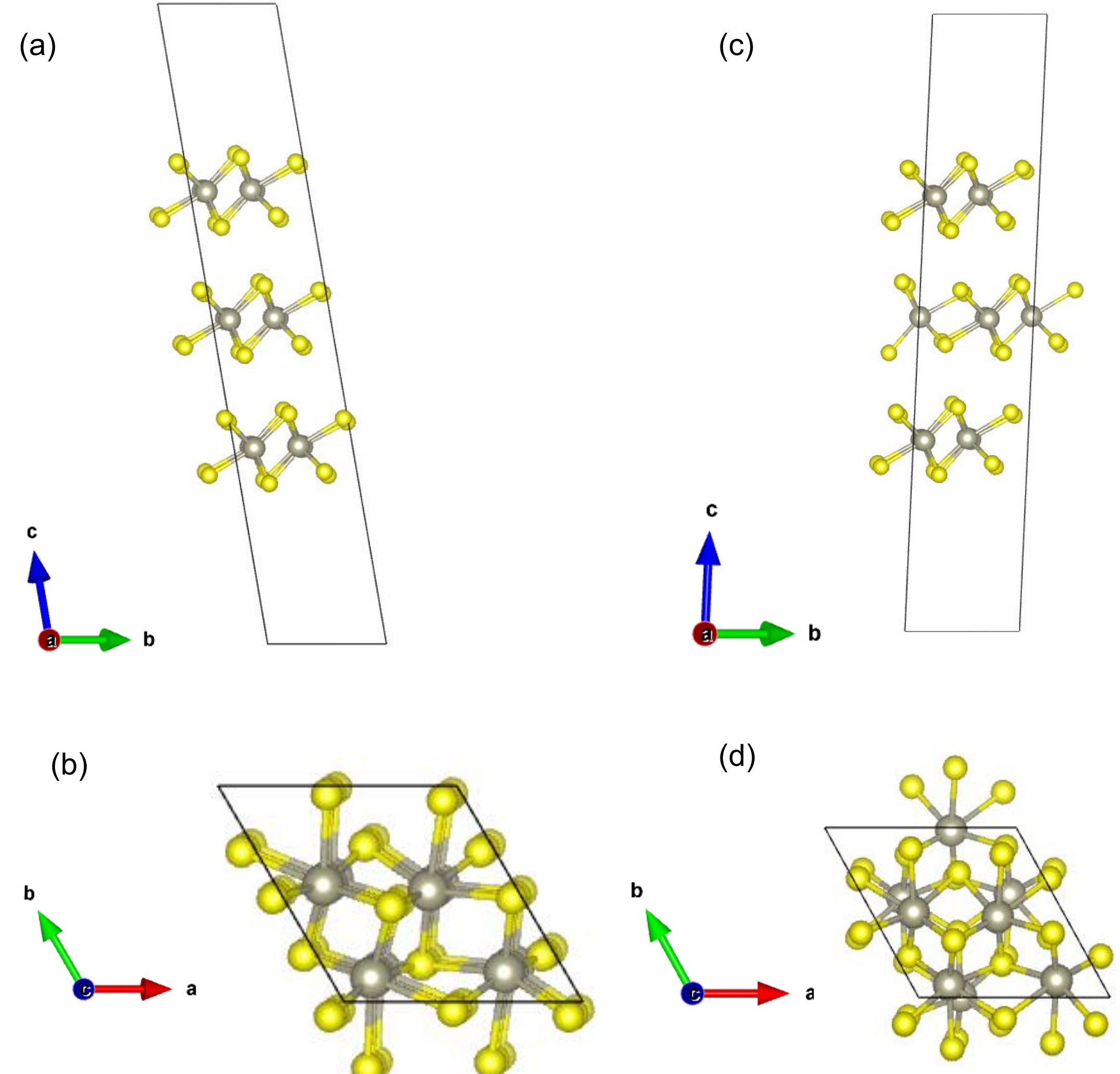

***Figure S12. (a,b)*** *Side and top views of AA stacked trilayer $ReS_2$.* ***(c,d)*** *Side and top views of AB stacked trilayer $ReS_2$.*

| | a | b | $\alpha$ | $\beta$ | $\gamma$ | $d_1$ | $d_2$ |
|---|---|---|---|---|---|---|---|
| AA | 6.38 | 6.48 | 91.55 | 103.98 | 118.84 | 3.10 | 3.08 |
| AB | 6.45 | 6.55 | 87.78 | 89.93 | 118.85 | 2.96 | 2.96 |

***Table S3.*** *The in-plane lattice parameters and the inter-layer distances in AA and AB stacked $ReS_2$. All distances are in units of Å. $d_1$ is the distance between the first and second layers. $d_2$ is the distance between the second and third layers.*

## Specific heat

The specific heat, entropy and free energy are calculated as a function of temperature, using the phonopy package. It can be observed that the specific heat is invariant of the stacking order.

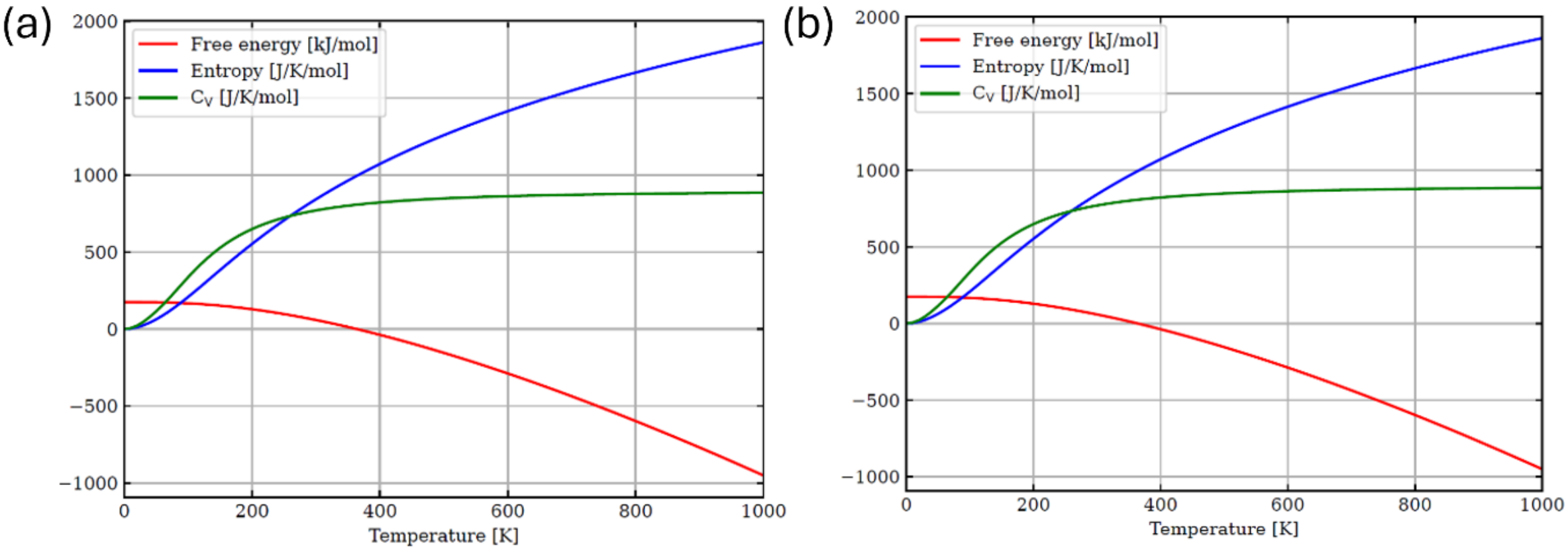

***Figure S13.*** *The specific heat, entropy, and free energy as a function of temperature for trilayer ReS2 with* ***(a)*** *AA stacking and* ***(b)*** *AB stacking.*